\documentclass{llncs}
\usepackage{verbdef}
\usepackage{amsmath}
\usepackage{amssymb}
\usepackage{mathrsfs}
\usepackage{mathtools}
\usepackage{mathabx}
\usepackage{amstext}
\usepackage{array}
\usepackage{xspace}
\usepackage{latexsym}
\usepackage{commands}
\usepackage[inference]{semantic}
\usepackage{stmaryrd}
\usepackage{booktabs}
\usepackage{tabularx}
\usepackage{multirow}
\usepackage{color}
\usepackage{listings}
\usepackage{paralist}
\usepackage[font=small]{caption}
\usepackage{subcaption}
\usepackage[table]{xcolor}
\usepackage{hyperref}
\usepackage{cleveref}
\usepackage{placeins}

\newcommand{\eqsep}{\rule{0.25\textwidth}{0.4pt}}

\newif\ifappendix
\appendixfalse

\newif\iftechrep
\techreptrue

\newcommand{\ra}[1]{\renewcommand{\arraystretch}{#1}}
\DeclareCaptionFormat{myformat}{\hrulefill\\\textbf{#1}#2#3}
\DeclareCaptionFormat{myformat2}{\\\textbf{#1}#2#3}
\crefname{figure}{Fig.}{Fig.}
\crefname{table}{Table}{Table}
\crefname{corollary}{corollary}{corollaries}
\Crefname{corollary}{Corollary}{Corollaries}
\crefname{section}{\S}{\S}
\crefname{appendix}{\S}{\S}

\MkLemma{wfctx}{
  If for every $\tybind{x}{\ty} \in \ctx$, $\iswellformed{\ctx}{\ty}$,
  then the only symbols that appear in $\denote{\ctx}$ are variables
  ($x,y$), arithmetic and equality symbols and uninterpreted
  functions, and no variable appears bound twice.
}
\MkLemma{wfheap}{
  If $\iswellformed{\ctx}{\heap}$, 
  then for each $\hplocty{\loc}{x}{\ty} \in \heap$:
\begin{enumerate}
  \item
    No binding $\tybind{x}{\ty'}$ appears in $\ctx$.
  \item
    $\denote{\heap}$ contains exactly one sub-assertion of the form $\sepmap{\loc}{x}$. 
\end{enumerate}
}
\MkLemma{schemawf}{
  If $\iswellformed{\ctx,\heap}{\funschema}$, 
  then the formal arguments or free variables of the output world of \funschema do not appear free in $\ctx$ or $\heap$.
}

\MkLemma{wfmeasures_old}{
  Suppose
  $\iswellformedMeas{}{\measDef{m}{\tybind{x}{\tyapp{\tycon}{\many\tvar}}}{e}}$,
  where
  \[\tydef{\tyapp{\tycon}{\tvar}}{\heap}{\evar}{\ty}.\]
  Assuming that $e$ is a well-typed term,
  \begin{align*}
  \mbox{If} &\quad&&
  s, h \vDash \exquant{\fv{\heap}}\denotef{\tybind{x}{\ty}}{\heap}
  \mbox{\quad and\quad }
  x_v \in \purety{\tycon}(A)\\
  \mbox{then} &\quad&& \denote{\fn{m}(x_v)}_{exp} s \in \mathsf{Values}
\end{align*}
}

\MkLemma{wfmeasures}{ If
  $\iswellformedMeas{}{\measDef{m}{\tybind{x}{\tyapp{\tycon}{\many\tvar}}}{e}}$,
  and $v \in \purety{\tycon}(A)$, then $\sepmeas{\sub{x}{v}e} \in
  \fn{Values}$.  }

\MkLemma{subtyping}{
  \label{lem:subtyping}
  \emphbf{[Subtyping]}
  \begin{thitemize}
    \item If \quad $\issubtype{\ctx}{}{\ty_1}{\ty_2}$
          \quad then \quad $\denote{\ctx} \Rightarrow \denote{\ty_1} \Rightarrow \denote{\ty_2}$
    \item If   \quad $\issubtype{\ctx}{}{\heap_1}{\heap_2}$ 
          \quad then \quad $\denote{\ctx} \Rightarrow \denote{\heap_1} \Rightarrow \denote{\heap_2}$ 
  \end{thitemize}
}

\MkLemma{heapsubtyping}{
If $\issubtype{\ctx}{}{\heap_1}{\heap_2}$ 
then for each $\loc \in (\dom{\heap_1} \cap \dom{\heap_2}) \cup \locs{\ctx}$,
  if $\heap_1 = \heap_1'\hpjoin\hplocty{\loc}{\evar}{\ty_1}$ and
  $\heap_2 = \heap_2'\hpjoin\hplocty{\loc}{\evar}{\ty_2}$, then
  \[
  \denote{\ctx} \Rightarrow \sprop{\ty_1}{x} \Rightarrow \sprop{\ty_2}{x}
  \]
}

\MkLemma{heapsubtypingdos}{
  If 
  \[
  \issubtype{\ctx}{}{\heap_1}{\heap_2}
  \]
  It then follows that
  \[
  \denotef{\Gamma}{\heap_1} \Rightarrow \denotef{\Gamma}{\heap_2}
  \]
}

\MkLemma{decomp}{
  Assume that $\Phi$, $\ctx$, and $\heap$ are (respectively) well
  formed global, local, and heap contexts. We may then write the denotation
  of these typing contexts in the following way:
  \begin{equation*}
    \denotef{\ctx}{\heap} = G \hpjoin H
  \end{equation*}
  where the following hold:
  \begin{enumerate}
    \item
      $G$ is \emph{pure}.
    \item $G$ is a conjunction of the assertions $$G_{\evar} \myeq 
      \sprop{\ty}{\evar}$$ for each $\tybind{\evar}{\ty} \in
      \ctx$. $G$ may thus be written $\cdots \wedge G_x \wedge
      \cdots$ for each such binding in $\ctx$.
    \item
      $H$ is a conjunction (via $\hpjoin$) of the assertions $$H_\loc
      \myeq (\loc = nil \vee \shprop{\ty}{\loc}{\evar})$$ for each
      $\hplocty{\loc}{\evar}{\ty} \in {\heap}$. $G$ may thus be
      written $\cdots H_\loc \cdots$ for each such binding in $\heap$.
  \end{enumerate}
}
\MkLemma{exprtypelemma}{
  \begin{flalign*}
    \mbox{If} & \quad \exprtype{\ctx}{\heap}{\expr}{\ty} &\\
    \mbox{Then} & \quad \denotef{\ctx}{\heap} = G \hpjoin H &
  \end{flalign*}
  such that
  \begin{enumerate}
    \item
      $G$ is \emph{pure}.
    \item
      $G \Rightarrow \sprop{\ty}{\expr}$
  \end{enumerate}
}
\MkLemma{ptrtype}{
  Suppose that $\exprtype{\ctx}{\heap}{x}{\tref{\loc}}$ and that
  $\iswellformed{\ctx}{}{\heap = \heap_0 \hpjoin \hplocty{\loc}{y}{\ty}}$. It
  then follows that
  $
  \denotef{\ctx}{\heap} = G \hpjoin H \hpjoin H_\loc
  $
  where $G$ is \emph{pure} and:
  \begin{enumerate}[(i)]
    \item
      $G \Rightarrow (x = \loc \wedge x \ne null)$
    \item
      $H = H' \hpjoin H_\loc \hpjoin H''$ such that
      \[G \hpjoin H_\loc \Rightarrow \shprop{\ty}{x}{y}\]
      and furthermore, if $\ty$ is a record type,
      \[G \hpjoin\shprop{\ty}{x}{y}\hpjoin H_\loc \Rightarrow \sepmap{x}{y}\]
      equivalently:
      \[
      G \hpjoin\shprop{\ty}{x}{y}\hpjoin H_\loc \Rightarrow \sepmap{x}{\tgenobj
        {\field_i}{\objfield{y}{\field_i}}}\]
  \end{enumerate}
}
\newcommand{\foldLoc}{\loc}
\newcommand{\foldExtra}[1]{\ensuremath{\hplocty{\foldLoc}{x}{#1}}}
\newcommand{\foldHyporhs}[3]{
  \exquant{\fv{#3}}
    \denote{\foldExtra{#2}\hpjoin #3} 
  \wedge
  \exquant{#1}
    \sepunfold{\foldLoc}{#1}
}
\newcommand{\foldHypolhs}[3]{\denotef{#1}{\foldExtra{#2}\hpjoin{#3}}}

\MkLemma{folding}{
  \emphbf{[Folding]}
  If $\heapfoldxy{\ctx}{\ty_1}{\heap_1}{\ty_2}{\heap_2}{x}{x}$ then
  $$\denote{\ctx} \Rightarrow 
         \denote{\hplocty{\loc}{x}{\ty_1} \hpjoin \heap_1} \Rightarrow 
         \exquant{\fv{\heap_2}}
         \denote{\hplocty{\loc}{x}{\ty_2} \hpjoin \heap_2}
         \wedge
        \exquant{y}\sepunfold{\loc}{y}$$
}

\MkLemma{pure}{
  Given the definition
  \[ \tydef{\tyapp{\tycon}{\tvar}}{\heap}{\evar}{\ty}, \] and the set of
  corresponding pure values, $\purety{\tycon}(A)$:
  \[
  \sepunfold{\loc}{x_f} \wedge \denote{\hplocty{\loc}{\evar}{\ty}\hpjoin{\heap}} \Rightarrow
  x_f \in \purety{\tycon}(A)
  \]
}
\MkLemma{returnlemma}{
  Suppose that
  \[
  \retstmttype{\ctx}{\heap\hpjoin\heap'}{s}
  \]
  and
  \[
  \ret{\equant{\many{\loc_\retId}}\tybind{\evar_\retId}{\ty_\retId}/\heap_\retId} \in \ctx \\
  \]
  It then follows that
  \[
  \htriplefun{\Phi}
          {\denotef{\ctx}{\heap\hpjoin\heap'}}
          {s}
          {\exquant{x_\retId}\denote{\tybind{x_\retId}{\ty_\retId}} \wedge \denote{\heap_\retId}}
  \]
}

\MkTheorem{procthm}{}{
  \emphbf{[Procedure Typing]} \\[4pt] 
  If \ $\pgmtype{\Phi}{\fun}{\funschema}$ \ then\
  $\htriplefun{\denote{\Phi}}{\Pre{\funschema}}{\Body{f}}{\Post{\funschema}}$.
}

\MkTheorem{procthmOLD}{Procedure Typing to Separation Logic}{
  \begin{flalign*}
    \mbox{If} &\quad&&
    \pgmtype{\Phi}{\fun}{\funschema}&\\
    \mbox{And} &\quad&&
    (P,Q,s) = \sfprop{\funschema}{f}&\\
    \mbox{Then}
    &\quad&&
      \htriplefun{\denote{\Phi}}{P}{f(\many{x})}{Q}&
  \end{flalign*}
}

\MkTheorem{stmtthm}{}{\emphbf{[Statement Typing]}
  \label{thm:stmtthm}
  \renderifthenlinenobrk
    {\stmttype{\ctx}{\heap}{\stmt}{\ctx'}{\heap'}}
    {\htriplefun{\denote{\Phi}}{\denotef{\ctx}{\heap}}{\stmt}{\denotef{\ctx'}{\heap'}}}
}

\MkTheorem{stmtprocthm}{}{\emphbf{[Typing Translation]}
  \begin{thitemize}
    \item \renderifthenlinenobrk
            {\stmttype{\ctx}{\heap}{\stmt}{\ctx'}{\heap'}}
            {\htriplefun{\denote{\Phi}}{\denotef{\ctx}{\heap}}{\stmt}{\denotef{\ctx'}{\heap'}}}
    \item \renderifthenlinenobrk
            {\pgmtype{\Phi}{\fun}{\funschema}} 
            {\htriplefun{\denote{\Phi}}{\Pre{\funschema}}{\Body{f}}{\Post{\funschema}}}
  \end{thitemize}
}

\MkTheorem{stmtthmOLD}{}{
  \emphbf{[Statement Typing to Separation Logic]}
  \begin{flalign*}
    \mbox{If}   & \quad && 
      \stmttype{\ctx}{\heap}{\stmt}{\ctx'}{\heap'}&\\
    \mbox{Let}  & \quad &&
    (P_f,Q_f,s_f) = \sfprop{\funschema_f}{f} 
    \mbox{\, for each } \tybind{f}{\funschema_f} \in \Phi&\\
    \mbox{Then} &\quad&&
    \inference{
      \htriplefun{\denote{\Phi}}{P_f}{f(\many{x})}{Q_f}&\\
      \vdots
    }{
      \htriplefun{\denote{\Phi}}
                 {\exquant{\many{V}}\denotef{\ctx}{\heap}}
                 {\stmt}
                 {\exquant{\many{V}}\denotef{\ctx'}{\heap'}}
    }&\\
    \mbox{Where} &\quad&&
    \many{V} = \fv{\ctx,\ctx',\heap,\heap'}&
    \end{flalign*}
}

\MkCorollary{typeproofs}{}{\emphbf{[Soundness]}
If \     $\stmttype{\varnothing}{\hpemp}{s}{\ctx}{\heap}$,\, 
 then\ $\htriplefun{\denote{\Phi}}{\ttrue}{s}{\ttrue}$
}

\MkCorollary{typeproofsOLD}{Separation Logic Proofs from Types}{
  \begin{flalign*}
    \mbox{If} &\quad&&
    \stmttype{\emptyset}{\hpemp}{s}{\ctx}{\heap}
    & \\
    \mbox{Then} &\quad&&
    \htriplefun{\denote{\Phi}}{\top}{s}{\top}&
\end{flalign*}
where no assertion fails and every memory access is safe.
}

\begin{document}

\setlength{\textfloatsep}{4pt plus 1.0pt minus 3.0pt}
\setlength{\abovedisplayskip}{4pt plus 1.0pt minus 2.0pt}
\setlength{\belowdisplayskip}{4pt plus 1.0pt minus 2.0pt}
\setlength{\belowdisplayshortskip}{4pt plus 1.0pt minus 2.0pt}

\newcounter{ex4}
\renewcommand{\theex4}{\alph{ex4}}

\setlength{\pdfpageheight}{\paperheight}
\setlength{\pdfpagewidth}{\paperwidth}




\title{Predicate Abstraction for~Linked~Data~Structures
}
\author{Alexander Bakst \and Ranjit Jhala}
\institute{University of California, San Diego\\
\email{\{abakst,jhala\}@cs.ucsd.edu}
}

\definecolor{gray_ulisses}{gray}{0.55}
\definecolor{castanho_ulisses}{rgb}{0.71,0.33,0.14}
\definecolor{preto_ulisses}{rgb}{0.41,0.20,0.04}
\definecolor{green_ulises}{rgb}{0.2,0.75,0}

\def\codesize{\small}

\definecolor{lightgray}{rgb}{.8,.8,.8}
\definecolor{darkgray}{rgb}{.4,.4,.4}
\definecolor{purple}{rgb}{0.65, 0.12, 0.82}
\definecolor{darkgreen}{RGB}{0,100,0}

\lstdefinelanguage{JavaScript}{
  morecomment=[l]{//},
  morecomment=[s]{/*}{*/},
  morestring=[b]',
  morestring=[b]",
  basicstyle=\codesize\ttfamily,
  keywordstyle=\color{blue},
  identifierstyle=\color{black},
  commentstyle=\color{darkgreen},
  stringstyle=\color{red},
  sensitive=true,
  emph={[1] 
    null},
  emphstyle={[1]\color{purple}},
  emph=
	{[2]
		void,int,bool,number,boolean,char,float,string,list
	},
  emph=
	{[3]
      break,case,catch,continue,debugger,default,do,ifnull,then,else,false,finally,for,function,
      if,in,instanceof,new,return,switch,this,throw,true,try,
      typeof,var,while,with,primitive,type
	},
  emphstyle={[3]\color{blue}},
  emph=
 	{[4]
      wind,unwind,conc,measure
 	},
  emphstyle={[4]\color{castanho_ulisses}\textbf}	
}

\lstdefinestyle{myjs}{
   language=JavaScript,
   columns=flexible,
   keepspaces=true,
   extendedchars=true,
   basicstyle=\codesize\ttfamily,
   showstringspaces=false,
   showspaces=false,
   numbers=none,                    
   numbersep=5pt,                   
   numberstyle=\tiny\color{red},    
   tabsize=2,
   breaklines=true,
   showtabs=false,
   captionpos=b
}

\lstnewenvironment{src}
{\lstset{language=JavaScript}}
{\lstset{style=myjs}}
{}

\lstMakeShortInline[language=JavaScript]@

\maketitle

\begin{abstract}
We present \emph{Alias Refinement Types} (\toolname),
a new approach that uses predicate-abstraction to
automate the verification of correctness properties
of linked data structures.
While there are many techniques for checking that a
heap-manipulating program adheres to its specification,
they often require that the programmer annotate the
behavior of each procedure, for example, in the form
of loop invariants and pre- and post-conditions.
We introduce a technique that lifts predicate abstraction
to the heap by factoring the analysis of data structures into two
orthogonal components:
(1)~Alias Types, which reason about the
\emph{physical} shape of heap structures, and
(2)~Refinement Types, which use simple predicates from an SMT
decidable theory to capture the \emph{logical} or semantic properties
of the structures.
%
We evaluate \toolname by implementing a tool that performs type
\emph{inference} for an imperative language, and empirically
show, using a suite of data-structure benchmarks, that \toolname
requires only 21\% of the annotations needed by other state-of-the-art
verification techniques.
\end{abstract}


\section{Introduction}\label{sec:intro}

Separation logic (SL)~\cite{Reynolds02} has proven
invaluable as a unifying framework for specifying and
verifying correctness properties of linked data structures.
Paradoxically, the richness of the logic has led to a problem
-- analyses built upon it are exclusively either expressive
\emph{or} automatic.
To \emph{automate} verification, we must restrict the logic to
decidable fragments, \eg list-segments~\cite{berdine2006smallfoot,LahiriQadeer08},
and design custom decision procedures~\cite{predator,haase2013seloger,BouajjaniSLAD2012,navarro2011separation,piskac2013automating}
or abstract interpretations~\cite{MagillTHOR08,cookCAV08,changPOPL08}.
Consequently, we lose expressiveness as the resulting
analyses cannot be extended to \emph{user}-defined structures.
To \emph{express} properties of user-defined structures, we must
fall back upon arbitrary SL predicates. We sacrifice
automation as we require programmer assistance to verify
entailments over such predicates~\cite{HTT,Chlipala11}.
Even when entailment is automated by specializing proof search,
the programmer has the onerous task of providing complex
auxiliary inductive invariants~\cite{ChinDNQ12,Qiu0SM13}.

We observe that the primary obstacle towards obtaining
expressiveness and automation is that in SL, machine
state is represented by monolithic assertions that conflate
reasoning about heap and data. While SL based tools commonly
describe machine state as a conjunction of a pure, heap
independent formula, and a \verb+*+ combination of heap
predicates, the heap predicates themselves conflate
reasoning about links (\eg reachability) and correctness
properties (\eg sizes or data invariants), which complicates
automatic checking and inference.

In this paper, we introduce \emph{Alias Refinement Types}
(\toolname), a subset of separation logic that reconciles
expressiveness and automation by \emph{factoring}
the representation of machine state along two independent
axes:
a \emph{``physical"} component describing the basic
shape and linkages between \emph{heap cells} and
a \emph{``logical"} component describing semantic
or relational properties of the \emph{data}
contained within them.
We connect the two components in order to describe
global logical properties and relationships
of heap structures, using \emph{heap binders}
that name pure ``snapshots'' of the mutable data
stored on the heap at any given point.

The separation between assertions about the heap's structure and
heap-oblivious assertions about pure values allow \toolname to
automatically \emph{infer} precise data invariants.
First, the program is type-checked with respect to the physical
type system.
Next, we generate a system of subtyping constraints over the \emph{logical}
component of the type system.
Because the logical component of each type is heap-oblivious, solving the
system of constraints amounts to solving a system of Horn clauses.
We use predicate abstraction to solve these constraints, thus yielding
precise refinements that summarize unbounded collections of objects.

In summary, this paper makes the following contributions:
\begin{itemize}
\item a description of \toolname and formalization of a constraint
  generation algorithm for inferring precise invariants of linked data
  structures;
\item a novel soundness argument in which types are interpreted as
  assertions in separation logic, and thus typing derivations are
  interpreted as proofs;
\item an evaluation of a prototype implementation that demonstrates
  \toolname is effective at verifying and, crucially, inferring data
  structure properties ranging from the sizes and sorted-ness of linked
  lists to the invariants defining binary search trees and red-black
  trees.
  Our experiments demonstrate that \toolname requires only 21\% of the
  annotation required by other techniques to verify intermediate
  functions in these benchmarks.
%
  \end{itemize}

\section{Overview} \label{sec:overview}


\begin{figure}[t]
\begin{minipage}{0.43\textwidth}
\rowcolors{2}{gray!30}{}
  \begin{center}
\ra{0.9}
{\small
\begin{tabular}{|m{0.53\textwidth}m{0.43\textwidth}|}
  \hline
  \multicolumn{2}{|c|}{\sfunty{abs}{\tint}{\tnat^1}}\\
  \hline
  \hline
@function abs(x){@ & $\tb{\x}{\tint}$          \\ 
@  if (0 <= x)   @  & $\grd{(0 \leq \x)};\tb{\x}{\tint}$\\ 
@    return x;   @  &  \\
@  var r = 0 - x;@  &
{$ \begin{aligned}
   &              \tb{\pvar{r}}{\refp{\vv = 0 - \x}}; \\
   &\grd{\lnot(0 \leq \x)} ; \tb{\x}{\tint}
   \end{aligned} $}
\\
@  return r;     @  & \\
@} @                & \\
\hline
\end{tabular}
}
\captionof{figure}{Refinement types}\label{fig:ex1}
\end{center}


\end{minipage}
\hfill
\begin{minipage}{0.55\textwidth}
\rowcolors{2}{gray!30}{}
  \begin{center}
\ra{0.9}
{\small
\begin{tabular}{|m{0.44\textwidth}m{0.55\textwidth}|}
\hline
\multicolumn{2}{|c|}{\small \sfunty{absR}
                            {\tb{\x}{\tobj{\fld{data}{\tint}}}}   
                            {()/\hb{x}{\tobj{\fld{data}{\tnat^2}}}}} \\
\hline
\hline
@function absR(x){@   & 
  $\ctx_0 \doteq \tb{\x}{\tref{\addr{\x}}}$\newline
  $\heap_0 \doteq \hb{\x}{\tobj{\fld{data}{\tint}}}$ \\ 
@  var d  = x.data;@  & $\ctx_1 \doteq \tb{\pvar{d}}{\tint};\ctx_0$ \\ 
@  var t  = abs(d);@  & $\ctx_2 \doteq \tb{\pvar{t}}{\tnat^1};\ctx_1$  \\
@  x.data = t;@       & $\heap_1 \doteq \hb{\x}{\tobj{\fld{data}{\vv = t}}}$\\
@  return;@           & \\
@}@                   & \\
\hline
\end{tabular}
}
\captionof{figure}{Strongly updating a location}\label{fig:ex2}
\end{center}


\end{minipage}
\end{figure}
\mypara{Refinements Types and Templates.}
A \emph{basic} refinement type is a basic type, \eg \tint, refined
with a formula from a decidable logic, \eg
%
$nat \doteq \reftp{\tint}{0 \leq \vv}$
is a refinement type denoting the set of non-negative integers,
where $\tint$ is the basic or \emph{physical} part of the type and the
refinement $0 \leq \vv$ is the \emph{logical} part.
A \emph{template} is a refinement type where, instead of concrete
formulas we have \emph{variables} $\kvar{}$ that denote the unknown
to-be-inferred refinements.
In the case that the refinement is simply \vtrue,
we omit the refinement (\eg  \mbox{\tint =\reftp{\tint}{\vtrue}}).
We specify the behaviors of functions using
refined function types:
$(\tb{x_1}{t_1},\ldots,\tb{x_n}{t_n}) \Rightarrow t$.
The input refinement types $t_i$ specify the function's \emph{preconditions} and
$t$ describes the \emph{postcondition}.

\mypara{Verification.}
\toolname splits verification into two phases:
(1)~\emph{constraint generation}, which traverses the program to
create a set of Horn clause constraints over the $\kvar{}$, and
(2)~\emph{constraint solving}, which uses an off the shelf
predicate abstraction based Horn clause solver \cite{LiquidPLDI08}
that computes a least fixpoint solution that yields refinement
types that verify the program. Here, we focus on the novel step~(1).

\mypara{Path Sensitive Environments.}
To generate constraints \toolname traverses the code, building up an
\emph{environment} of type bindings, mapping program variables to
their refinement types (or templates, when the types are unknown.)  At
each call-site (resp. return), \toolname generates constraints that
the arguments (resp. return value) are a subtype of the input
(resp. output) type.
Consider @abs@ in \cref{fig:ex1} which computes the absolute
value of the integer input @x@.
\toolname creates a template $(\tint) \Rightarrow \reftp{\tint}{\kvar{1}}$
where $\kvar{1}$ denotes the unknown output refinement. (We write $\tnat^1$
in the figure to connect the inferred refinement with its $\kvar{}$.)
In \cref{fig:ex1}, the environment after each statement
is shown on the right side.
The initial environment contains a binder for @x@, which
assumes that \x may be \emph{any} $\tint$.
In each branch of the @if@ statement, the environment
is extended with a \emph{guard} predicate reflecting
the condition under which the branch is executed.
As the type \reftp{\tint}{\vv = \x} is problematic
if @x@ is mutable,
we use SSA renaming to ensure each variable is assigned (statically)
at most once.

\mypara{Subtyping.}
%
The returns in the \emph{then} and \emph{else} yield subtyping constraints:
\begin{align}
 \begin{split}
\tb{\x}{\tint}, 0 \leq \x
 \ \vdash\ & \reftp{\tint}{\vv = \x}
 \ \relst\ \reftp{\tint}{\kvar{1}} \\
\tb{\x}{\tint}, \lnot(0 \leq \x), \tb{\pvar{r}}{\reftp{\tint}{\vv = 0 - \x}}
 \ \vdash\ & \reftp{\tint}{\vv = \pvar{r}}
 \ \relst\ \reftp{\tint}{\kvar{1}}
 \label{eqn:abs1}
 \end{split}
 \intertext{which respectively reduce to the Horn implications}
 (\vtrue \wedge 0 \leq \x)
 \ \Rightarrow\ & (\vv = \x)
 \ \Rightarrow\ \kvar{1}\nonumber \\
 (\vtrue \wedge \lnot(0 \leq \x) \wedge \pvar{r} = 0 - \x)
 \ \Rightarrow\ & (\vv = \pvar{r})
 \ \Rightarrow\ \kvar{1}\nonumber
 \end{align}
By predicate abstraction \cite{LiquidPLDI08} we find the solution
${\kvar{1} \doteq 0 \leq \vv}$ and hence infer that the returned
value is a $\tnat$, \ie non-negative.

\rowcolors{2}{gray!30}{}
\begin{figure*}[t] 
\ra{0.5}
  \begin{center}
\begin{tabular}{|m{0.25\textwidth}m{0.72\textwidth}|}
  \hline
  \multicolumn{2}{|c|}{\sfunty{absL}
                              {\tb{\x}{\tlist{\tint}}}
                              {\tvoid / \hb{\x}{\tlist{\tnat^3}}}} \\
  \hline
  \hline
@function absL(x){ @&
 $\ctx_0 \doteq \tb{\x}{\tref{\addr{\x}}}$,
 $\heap_0 \doteq  \hb{\x}{\tb{x_0}{\tlist{\tint}}}$
 \\ 
@  //: unfold(&x); @ &
    $\ctx_1 \doteq \ctx_0$ \newline
    $\heap_1 \doteq \hb{\x}{\tb{x_1}{\tobj{\fld{data}{\tint},\fld{next}{\trefm{\addr{t}}}}}} \hsep \hb{t}{\tb{t_0}{\tlist{\tint}}}$
\\
@  var d  = x.data;@\newline
@  x.data = abs(d);@\newline
@  var xn = x.next;@ &
    $\ctx_2 \doteq \tb{\pvar{d}}{\tint},\, 
\tb{\pvar{xn}}{\reftp{\trefm{\addr{t}}}{\vv = \pfield{x_2}{next}}},\, \ctx_1$
    \newline
    $\heap_2 \doteq 
\hb{\x}{\tb{x_2}{\tobj{\fld{data}{\tnat^1},\fld{next}{\trefm{\addr{t}}}}}}
\hsep \hb{t}{\tb{t_0}{\tlist{\tint}}}$
\\
@  if (xn == null){@ & \\
@    //: fold(&x); @ & 
$\ctx_3 \doteq \isnull{\pvar{xn}},\, \ctx_2$, $\heap_3 \doteq\hb{\x}{\tlist{\tnat^3}} $\\
@    return;       @  & \\
@  }               @  & \\
@  absL(xn);       @  & 
      $\ctx_4 \doteq \isnonnull{\pvar{xn}};\ctx_2$\newline
      $\heap_4 \doteq \hb{\x}{\tb{x_2}{\tobj{\fld{data}{\tnat^1},\fld{next}{\trefm{\addr{t}}}}}} \hsep \hb{t}{\tb{t_1}{\tlist{\tnat^3}}}$
\\
@  //: fold(&x);   @  &\cellcolor[gray]{1.0} $\ctx_5 \doteq \ctx_4$ $\heap_5 \doteq \hb{\x}{\tb{x_3}{\tlist{\tnat^3}}}$ \\ 
@  return;         @  & \\
@}                 @  & \\
\hline
\end{tabular}
\caption{Strongly updating a collection. The \code{fold} and \code{unfold} annotations are
automatically inserted by a pre-analysis \iftechrep (\cref{sec:annots}) \else \cite{arttechrep}\fi}\label{fig:ex3}
\end{center}
\end{figure*}


\mypara{References and Heaps.}
In \cref{fig:ex2}, @absR@ takes a \emph{reference} to
a structure containing an $\tint$ valued @data@ field,
and updates the @data@ field to its @abs@olute value.
We use \kvar{2} for the output refinement; hence
the type of @absR@ desugars to:
$\funty{\tb{\x}{\addr{\x}}}
       {\hb{\x}{\tobj{\fld{data}{\tint}}}}
       {\tvoid}
       {\hb{\x}{\tobj{\fld{data}{\kvar{2}}}}}$
which states that @absR@ \emph{requires} a parameter @x@ that
is a \emph{reference} to a \emph{location} named @&x@.
in an \emph{input heap} where @&x@ contains a structure with an
\tint-valued @data@ field.
%
%
The function returns \tvoid (\ie no value) in an \emph{output heap}
where the location @&x@ is \emph{updated} to a structure with a \kvar{2}-valued
@data@-field.

We extend the constraint generation to precisely track updates to
locations. In \cref{fig:ex2}, each statement of the code is followed
by the environment $\ctx$ and heap \heap that exists after the
statement executes.
Thus, at the start of the function, @x@ refers to a
location, @&x@, whose @data@ field is an arbitrary \tint.
The call @abs(d)@ returns a $\kvar{1}$ that is bound to
@t@, which is then used to \emph{strongly update} the
@data@ field of @&x@ from \tint to $\kvar{1}$.
At the \returnName we generate a constraint that the return
value and heap are sub-types of the function's return type
and heap. Here, we get the \emph{heap subtyping} constraint:
\begin{align*}
  \tb{\x}{\tref{\addr{\x}}},\ \tb{\pvar{d}}{\tint},\ \tb{\pvar{t}}{\kvar{1}}
  \ \vdash\ & \  \hb{\x}{\tobj{\fld{data}{\vv = \pvar{t}}}} \relst\ \hb{\x}{\tobj{\fld{data}{\kvar{2}}}}
\end{align*}
which reduces by field subtyping to the implication:
${\kvar{1} \sub{\vv}{\pvar{t}}\Rightarrow (\vv = \pvar{t}) \Rightarrow \kvar{2}}$
which (together with the previous constraints) can be solved to
$\kvar{2} \doteq 0 \leq \vv$ letting us infer that @absR@
updates the structure to make @data@ non-negative.
This is possible because the \kvar{} variables denote pure formulas,
as reasoning about the heap shape is handled by the alias type system.
Next we see how this idea extends to infer strong updates to
collections of linked data structures.

\mypara{Linked Lists.} Linked lists can be described as iso-recursive
alias types \cite{AliasTypesRec}. The definition
\[
\mathtt{type}\ \tlist{A}\ \doteq\
    \exists! \loc \mapsto \tb{t}{\tlist{A}}.
       \tb{h}{\tobj{\fld{data}{A}, \fld{next}{\trefm{\loc}}}}
\]
says \tlist{A} is a \emph{head} structure with a @data@ field
of type $A$, and a @next@ field that is either \vnull or a
reference to the \emph{tail}, denoted by the \trefm{\loc} type.
The heap $\loc \mapsto \tb{t}{\tlist{A}}$ denotes
that a \emph{singleton} \tlist{A} is stored at the location
denoted by $\loc$ \emph{if} it is reachable at runtime.
The $\exists!$ quantification means that the tail is
\emph{distinct} from every other location, ensuring
that the list is inductively defined.

 Consider @absL@ from \cref{fig:ex3}, which updates each @data@
field of a list with its @abs@olute value. As before, we start
by creating a $\kvar{3}$ for the unknown output refinement, so
the function gets the template
$$
\funty{\tb{x}{\tref{\addr{x}}}}
      {\hb{x}\tb{x_0}{\tlist{\tint}}}
      {\tvoid}
      {\hb{x}\tb{x_{r}}{\tlist{\kvar{3}}}}
$$
\Cref{fig:ex3} shows the resulting environment and heap after
each statement.

The annotations @unfold@ and @fold@ allow \toolname
to manage updates to collections such as lists.
In \toolname, the user \emph{does not} write @fold@ and @unfold@
annotations; these may be inferred by a straightforward analysis of
the program\iftechrep (\cref{sec:annots}) \else \cite{arttechrep} \fi.


\mypara{Unfold.}
The location \addr{\x} that the  variable @x@ refers to initially
contains a $\tlist{\tint}$ named with a \emph{heap binder} $x_0$.
The binder $x_0$ may be used in refinements.
Suppose that @x@ is a reference to a location containing a value of
type \tlist{A}.
We require that before the fields of @x@ can be accessed, the
list must be \emph{unfolded} into a head cell and a
tail list.
This is formalized with an @unfold(&x)@ operation that
unfolds the list at $\addr{\x}$ from $\hb{\x}\tb{x_0}{\tlist{\tint}}$ to
$$
\hb{\x}\tb{x_1}{\tobj{\fld{data}{\tint},\fld{next}{\trefm{\addr{t}}}}}
\hsep
\hb{t}\tb{t_0}{\tlist{\tint}},
$$
corresponding to \emph{materializing} in shape analysis.
The type system guarantees that the head structure and
(if @next@ is not \vnull) the newly unfolded tail structure
are unique and distinct.
So, after unfolding, the structure at $\addr{\x}$ can
be strongly updated as in @absR@.  Hence, the field assignment
generates a fresh binder $x_2$ for the updated structure
whose @data@ field is a \kvar{1}, the output of @abs@.

\mypara{Fold.} After updating the @data@ field of the head, the
function tests whether the @next@ field assigned to @xn@ is \vnull,
and if so returns.
Since the expected output is a list, \toolname requires that
we \emph{fold} the structure back into a \tlist{\kvar{3}} -- effectively
computing a \emph{summary} of the structure rooted at $\addr{\x}$.
As @xn@ is \vnull and
$\tb{\pvar{xn}}{\reftp{\trefm{\addr{t}}}{\vv = \pfield{x_2}{next}}}$,
@fold(&x)@ converts
$\hb{\x}{\tb{x_2}{\tobj{\fld{data}{\kvar{1}},\fld{next}{\trefm{\addr{t}}}}}}$
to
${\hb{\x}{\tlist{\kvar{3}}}}$
\emph{after} generating a heap subtyping constraint which forces the ``head''
structure to be a subtype of the folded list's ``head'' structure.
\begin{equation}
\ctx_3\ \vdash\
   \hb{\x}{\tb{x_2}{\tobj{\fld{data}{\kvar{1}},\ldots}}}
   \ \relst\
   \hb{\x}{\tb{x_2}{\tobj{\fld{data}{\kvar{3}},\ldots}}}
   \label{eqn:absL1}
\end{equation}
If instead, @xn@ is non-null, the function updates the tail by
recursively invoking @absL(xn)@. In this case, we can inductively
assume the specification for @absL@ and so in the heap \emph{after} the recursive call,
the tail location \addr{t} contains a $\tlist{\kvar{3}}$.
As @xn@ and hence the @next@ field of $x_2$ is non-null,
the @fold(&x)@ transforms
$$
\hb{\x}{\tb{x_2}{\tobj{\fld{data}{\kvar{1}},\fld{next}{\trefm{\addr{t}}}}}}
\hsep
\hb{t}{\tb{t_1}{\tlist{\kvar{3}}}}
$$
into
$\hb{x}{\tlist{\kvar{3}}}$, as required at the \returnName, by
generating a heap subtyping constraints for the head and tail:
\begin{align}
\ctx_5 \ \vdash \ & \hb{\x}{\tb{x_2}{\tobj{\fld{data}{\kvar{1}},\ldots}}}
                  \ \relst\
                  \hb{\x}{\tb{x_2}{\tobj{\fld{data}{\kvar{3}},\ldots}}}
                  \label{eqn:absL2} \\
\ctx_5\ \vdash\ &
                  \hb{t}{\tb{t_1}{\tlist{\kvar{3}}}}
                  \ \relst\
                  \hb{t}{\tb{t_1}{\tlist{\kvar{3}}}}
                  \label{eqn:absL3}
\end{align}
The constraints \cref{eqn:absL1}, \cref{eqn:absL2} and
\cref{eqn:absL3} are simplified field-wise into the implications
$\kvar{1} \Rightarrow \kvar{3}$, $\kvar{1} \Rightarrow \kvar{3}$ and
$\kvar{3} \Rightarrow \kvar{3}$ which, together with the previous
constraints (\cref{eqn:abs1}) solve to: $\kvar{3} \doteq 0 \leq \vv$.
Plugging this back into the template for @absL@ we see that we have
automatically inferred that the function \emph{strongly} updates the
contents of the input list to make \emph{all} the @data@ fields
non-negative.

\toolname \emph{infers} the update the type of the value
stored at @&x@ at @fold@ and @unfold@ locations because reasoning
about the shape of the updated list is delegated to the alias type
system.
Prior work in refinement type inference for imperative
programs~\cite{LiquidPOPL10} can not type check this simple example as
the physical type system is not expressive enough.
Increasing the expressiveness of the \emph{physical} type system
allows \toolname to ``lift'' invariant inference to collections of
objects.

\rowcolors{2}{gray!30}{}
\begin{figure*}[t!]
\ra{0.5}
\setlength{\tabcolsep}{0.00em}
\begin{tabular}{|rm{0.34\textwidth}m{0.66\textwidth}|}
  \hline
  \multicolumn{3}{|c|}{\sfunty{insert}
                              {A, \tb{\x}{?\tlist{A}}}
                              {\reftp{\tlist{A}}{(\mlen{\vv} = 1 + \mlen{x})^4}}} \\
  \hline
  \hline
&@function insert(k, x)@ & 
  $\ctx_0 = \tb{\pvar{k}}{A};\ \tb{\x}{\trefm{\addr{\x}}}$
  $\heap_0 = \hb{\x}{\tb{x_0}{\tlist{A}}}$ \\ 
&@  if (x == null) {    @ & 
\\
\iftechrep{\scriptsize{\pgmpoint{A}:}}\else{}\fi
&@    var y =           @ \newline
 @     {data:k,next:null};@ &
$\ctx_1 \doteq \tb{\y}{\tref{\addr{\y}}};\isnull{\x};\ctx_0$\newline
$\heap_1 \doteq \hb{\x}{\tb{x_0}{\tlist{A}}}
 \hsep \hb{\y}{\tb{y_0}{\tobj{\fld{data}{A}, \fld{next}{\tnull}}}}$\\
\iftechrep{\scriptsize{\pgmpoint{B}:}}\else{}\fi
&@    //: fold(&y)      @& $\ctx_2 \doteq \mlen{y_1} = 1;\ctx_1$\newline
                          $\heap_2 \doteq
                          \hb{\x}{\tb{x_0}{\tlist{A}}}
                            \hsep \hb{\y}{\tb{y_1}{\tlist{A}}}$ \\
\iftechrep{\scriptsize{\pgmpoint{C}:}}\else{}\fi
&@    return y;         @ \newline
 @  }                   @ & \\
&@  //: unfold(&x)      @ & 
$\ctx_3 \doteq \mlen{x_0} = 1 + \mlen{t_0};\isnonnull{x};\ctx_0 $\newline
$\heap_3 \doteq \hb{\x}{\tb{x_1}{\tobj{\fld{data}{a}, \fld{next}{\trefm{\addr{t}}}}}}
                            \hsep \hb{t}{\tb{t_0}{\tlist{a}}}$ \\
&@  if (k <= x.data) {  @ & \\
&@    var y =          @ \newline
 @     {data:k,next:x};@ &
$\ctx_4 \doteq \tb{\y}{\tref{\addr{\y}}};\ctx_3$\newline
$\heap_4 \doteq \hb{\y}{\tb{y_2}{\tobj{\fld{data}{A}, \fld{next}{\trefm{\addr{x}}}}}} \hsep \heap_3$ \\
&@    //: fold(&x)      @ &
  $\ctx_5 \doteq \mlen{x_2} = 1 + \mlen{t_0};\ctx_4$\newline
  {
    $\heap_5 \doteq  \hb{\x}{\tb{x_2}{\tlist{A}}} 
              \hsep \hb{\y}{\tb{y_2}{\tobj{\fld{data}{A}, \fld{next}{\trefm{\addr{x}}}}}}
    $}\\
&@    //: fold(&y)      @ &
  $\ctx_6 \doteq \mlen{y_3} = 1 + \mlen{x_2};\ctx_5$
  $\heap_6 \doteq  \hb{\x}{\tb{y_3}{\tlist{A}}}$\\
&@    return y;         @ \newline
 @  }                   @ & 
\\
&@  var z  = x.next;    @ \newline
 @  var u  = insert(k,z);@ \newline
 @  x.next = u;         @ &  
$\ctx_7 \doteq \tb{u_0}{\kvar{4} \sub{x_0}{t_0}};\tb{\pvar{u}}{\tref{\addr{u}}};\tb{\pvar{z}}{\trefm{\addr{t}}};\ctx_3$\newline
$\heap_7 \doteq
\hb{\x}{\tb{x_1}{\tobj{\fld{data}{A}, \fld{next}{\tref{\addr{u}}}}}}
                             \hsep \hb{\pvar{u}}{\tb{u_0}{\tlist{A}}}$
\\
&@  //: fold(&x)        @ &
$\ctx_8 \doteq \mlen{x_2} = 1 + \mlen{u_0};\ctx_7$
$\heap_8 \doteq \hb{\x}{\tb{x_2}{\tlist{A}}}$\\
&@  return x;           @ \newline
 @}                     @ & \\
\hline
\end{tabular}
\caption{Inserting into a collection}\label{fig:ex4}
\end{figure*}


\mypara{Snapshots.}
So far, our strategy is to factor reasoning about pointers and the heap
into a ``physical'' alias type system, and functional properties
(\eg values of the @data@ field) into quantifier- and heap-free ``logical''
refinements that may be inferred by classical predicate abstraction.
However, reasoning about recursively defined properties, such as the
length of a list, depends on the interaction between the physical and
logical systems.

We solve this problem by associating recursively defined properties
\emph{not} directly with mutable collections on the heap, but with
immutable \emph{snapshot values} that capture the contents of the
collection at a particular point in time.
These snapshots are related to the sequences of pure values that
appear in the definition of predicates such as \fn{list} in
\cite{Reynolds02}.
Consider the heap $\heap$ defined as:
\[
\addr{x_0} \mapsto \tb{h}{\tobj{\flde{data}{0}, \flde{next}{\addr{x_1}}}} \hsep
\addr{x_1} \mapsto \tb{t}{\tobj{\flde{data}{1}, \flde{next}{\vnull}}}
\]
We say that \emph{snapshot of \addr{x_0} in \heap} is
the value $v_0$ defined as:
\[
v_0 \doteq (\addr{x_0}, \tobj{\flde{data}{0}, \flde{next}{v_1}}) \quad
v_1 \doteq (\addr{x_1}, \tobj{\flde{data}{1}, \flde{next}{\vnull}})
\]

Now, the logical system can avoid reasoning about the heap
reachable from $x_0$ -- which depends on the heap --
and can instead reason about the length of the snapshot $v_0$
which is independent of the heap.

\mypara{Heap Binders.}
We use \emph{heap binders} to name snapshots in the refinement logic.
In the desugared signature for @absR@ from \cref{fig:ex2},
$$
\funty{\tb{x}{\tref{\addr{x}}}}
      {\hb{x}\tb{x_0}{\tlist{\tint}}}
      {\tvoid}
      {\hb{x}\tb{x_{r}}{\tlist{\tnat}}}
$$
the name $x_0$ refers to the snapshot of input heap at \addr{x}.
In \toolname, no reachable cell of a folded recursive structure (\eg
the list rooted at \addr{x}) can be \emph{modified} without first
\emph{unfolding} the data structure starting at the root: references
pointing into the cells of a folded structure may not be dereferenced.
Thus we can soundly update heap binders \emph{locally} without updating
transitively reachable cells.
%
%

\mypara{Measures.}
We formalize structural properties like the \emph{length} of a list or
the \emph{height} of a tree and so on, with a class of
recursive functions called \emph{measures}, which are catamorphisms
over (snapshot values of) the recursive type.  For example, we specify
the length of a list with the measure:
\begin{align*}
& \fn{len} :: \tlist{A} \Rightarrow \tint
&&
\fn{len}(\vnull) = 0 &&
\fn{len}(x) = 1 + \fn{len}(x.next)
\end{align*}
We must reason \emph{algorithmically} about these recursively defined
functions.
The direct approach of encoding measures as \emph{background axioms}
is problematic due to the well known limitations and brittleness of
quantifier instantiation heuristics~\cite{simplifyj}.
Instead, we encode measures as uninterpreted functions, obeying the
congruence axiom,
$\forall x, y. x = y \Rightarrow f(x) = f(y)$.
Second, we recover the semantics of the function by adding
\emph{instantiation constraints} describing the measure's semantics.
We add the instantiation constraints at @fold@ and @unfold@
operations, automating the reasoning about measures while retaining
completeness \cite{KuncakPOPL10}.

Consider @insert@ in \cref{fig:ex4}, which adds a key
@k@ of type @A@ into its position in an (ordered) \tlist{A},
by traversing the list, and mutating its links to accomodate
the new structure containing @k@.
We generate a fresh $\kvar{4}$ for the output type
to obtain the function template:
\[
(A, \tb{\x}{\trefm{\addr{\x}}})/ {\hb{\x}{\tb{x_0}{\tlist{A}}}}
\Rightarrow {\tref{\addr{l}}}/ {\hb{l}{\reftp{\tlist{A}}{\kvar{4}}}}
\]
Here, the snapshot of the input list \x\xspace upon entry is named
with the heap binder $x_0$; the output list must satisfy the (as yet unknown)
refinement $\kvar{4}$.

Constraint generation proceeds by additionally instantiating measures
at each @fold@ and @unfold@.
When @x@ is null, the @fold(&y)@ transforms the binding
$\hb{\y}{\tb{y_0}{\tobj{\fld{data}{A}, \fld{next}{\tnull}}}}$
into a (singleton) list
${\hb{\y}{\tb{y_1}{\tlist{A}}}}$ and so we add
the instantiation constraint ${\mlen{y_1} = 1}$
to the environment. Hence, the subsequent @return@ yields a subtyping
constraint over the output list that simplifies to the implication:
\begin{equation}
  \mlen{x_0} = 0 \wedge \mlen{y_1} = 1 \Rightarrow \vv = y_1 \Rightarrow \kvar{4}
  \label{eqn:ex4A}
\end{equation}
When @x@ is non-null, @unfold(&x)@ transforms the binding ${\hb{\x}{\tb{x_0}{\tlist{A}}}}$ to
$$    \hb{\x}{\tb{x_1}{\tobj{\fld{data}{a}, \fld{next}{\trefm{\addr{t}}}}}}
\hsep \hb{t}{\tb{t_0}{\tlist{A}}}
$$
yielding the instantiation constraint ${\mlen{x_0} = 1 + \mlen{t_0}}$
that relates the length of the list's snapshot with that of its tail's.
When @k <= x.data@ the subsequent folds create the binders $x_2$ and $y_3$
with instantiation constraints relating their sizes. Thus, at the @return@
we get the implication:
\begin{equation}
  \mlen{x_0} = 1 + \mlen{t_0} \wedge \mlen{x_2} = 1 + \mlen{t_0} \wedge \mlen{y_3} = 1 + \mlen{x_2} \Rightarrow \vv = y_3 \Rightarrow \kvar{4}
  \label{eqn:ex4B}
\end{equation}
Finally, in the else branch, after the recursive call to @insert@, and subsequent fold, we get the
subtyping implication
\begin{equation}
  \mlen{x_0} = 1 + \mlen{t_0} \wedge \kvar{4} \sub{u_0,t_0}{\vv,x_0} \wedge \mlen{x_2} = 1 + \mlen{u_0} \Rightarrow \vv = x_2 \Rightarrow \kvar{4}
  \label{eqn:ex4C}
\end{equation}
The recursive call that returns $u_0$ constrains it to satisfy
the unknown refinement \kvar{4} (after substituting $t_0$ for the
input binder $x_0$). Since the heap is factored out by the type
system, the classical predicate abstraction fixpoint computation
solves \cref{eqn:ex4A,eqn:ex4B,eqn:ex4C} to $\kvar{4} \doteq
\mlen{\vv} = 1 + \mlen{x_0}$ inferring a signature that
states that @insert@'s output has size one more than the input.




\begin{figure}[t!]
\ra{0.8}
  \begin{center}
\begin{tabular}{|l|}
  \hline
  \multicolumn{1}{|c|}{\color{darkgreen}{$\mathtt{(\tb{\x}{?\tlist{A}}) \Rightarrow
    \reftp{?{incList}[A]}{\mlen{\vv}=\mlen{\x}}}$}}\\
  \hline
  \hline
@function insertSort(x){      @ \\
@  if (x == null) return null;@ \\
@  //: unfold(&x);            @ \\
@  var y = insertSort(x.next);@ \\
@  var t = insert(x.data, y); @ \\
@  //: fold(&t);              @ \\
@  return t;                  @ \\
@}                            @ \\
  \hline
\end{tabular}
\caption{Insertion Sort}\label{fig:ex5}
\end{center}
\end{figure}

\mypara{Abstract Refinements.}
Many important invariants of linked structures require us
to reason about relationships \emph{between} elements of the
structure.
Next, we show how our implementation of \toolname
allows us to use \emph{abstract refinements}, developed in the purely
functional setting \cite{vazou13}, to
verify relationships between elements of linked
data structures, allowing us to prove that @insertSort@ in
\cref{fig:ex5} returns an ordered list.
%
To this end, we parameterize types with \emph{abstract refinements}
that describe relationships between elements of the structure. For example,
\rowcolors{1}{}{}
$$
\mathtt{type}\ \tlist{A}\tref{p}\ \doteq
\exists! l \mapsto \tb{t}{\tlist{\reftp{A}{p(\mathtt{data}, \vv)}}\tref{p}}.
     \tb{h}{\tobj{\fld{data}{A}, \fld{next}{\trefm{l}}}}
$$
is the list type as before, but now parameterized by
an abstract refinement $p$ which is effectively a
relation between two $A$ values.
The type definition  states that, if
the data fields have values $x_1,\ldots,x_n$
where $x_i$ is the $i^{th}$ element of the list, then
\emph{for each} $i < j$ we have $p(x_i, x_j)$.

\mypara{Ordered Lists.}
We \emph{instantiate} the refinement parameters with concrete refinements
to obtain invariants about linked data structures. For example,
increasing lists are described by the type $incList[A] \doteq \tlist{A}\tref{(\leq)}$.


\mypara{Verification.}
Properties like sortedness
may be \emph{automatically infered}
by using liquid typing \cite{LiquidPLDI08}.
\toolname infers the types:
\begin{align*}
\mathtt{insertSort} :: (?\tlist{A}) \Rightarrow \tolist{A} &&
\mathtt{insert}:: (A, ?\tolist{A}) \Rightarrow \tolist{A}
\end{align*}
\ie that @insert@ and @insertSort@ return sorted lists.
Thus, alias refinement types, measures, and abstract refinements
enable both the specification and automated verification of
functional correctness invariants of linked data structures.

 \rowcolors{1}{}{}
\section{Type Inference}\label{sec:cgen}
\begin{figure}[t!]
{\small
  \ra{0.9}
  \begin{tabular}{l >{$}l<{$} >{$}l<{$}}
\emph{\textbf{Expressions}} & \expr ::= &
                   \vnum 
           \spmid  \vtrue \spmid \vfalse
           \spmid \vnull
           \spmid  \vref_\loc
           \spmid  \evar
           \spmid  \expr \op \expr \\[0.01in]
\emph{\textbf{Statements}} & \stmt ::= &
                       \stmt; \stmt 
            \spmid      \evar = \expr
            \spmid      y = \access{\evar}{\field}
            \spmid      \access{\evar}{\field} \ifappendix =_z \else = \fi \expr
            \spmid      \ite{\expr}{\stmt}{\stmt}
            \\
           &&
            \spmid      \return{\expr}
           \spmid     \evar \ifappendix =_z \allocl{\loc}{\many{\field:\expr}} 
                              \else = \alloc{\many{\field:\expr}} \fi 
            \spmid      \evar = \funcall{\many{\expr}}
            \\
           &&
            \spmid      \ifappendix \unwindBind{\loc}{\many{y}} \else \unwind{\loc} \fi
            \spmid      \ifappendix \windBind{\loc}{x} \else \wind{\loc} \fi
            \spmid      \ifappendix \concBind{x}{y} \else \conc{x} \fi
          \ifappendix
           \spmid     \pad{\loc}{x}
          \else
           \spmid     \padnone{\loc}
          \fi
\\
\emph{\textbf{Programs}} & \prog ::= & \many{\fundecl{f}{\many{\evar}}}\\
\end{tabular}
}%

{\small
\ra{0.9}
\begin{tabular}{l>{\ } >{$}l<{$} >{$}l<{$} >{$}l<{$}}
\emph{\textbf{Primitive Types}} & \tprim ::= & 
                    \tint
            \spmid  \tbool
            \spmid  \tvar
            \spmid  \tnull
            \spmid  \tref{\loc}
            \spmid  \trefm{\loc}
            \\[0.01in]
\emph{\textbf{Types}} & \tbase ::= &
                  \tprim
           \spmid \tyapp{\tycon}{\many{\ty}}
           \spmid \tgenobj{\field}{\ty}\\[0.01in]
\emph{\textbf{Refined Types}} & \ty ::= & \reft\\[0.01in]
\emph{\textbf{Type Definition}} & \tycon ::=    & 
                      \tydef{\tyapp{\tycon}{\many{\tvar}}}
                            {\heap}
                            {\evar}
                            {\tobj{\many{f:\ty}}}\\[0.01in]
\emph{\textbf{Contexts}} & \ctx ::= & \emptyset \spmid \tybind{\evar}{\ty};\ctx 
                                \spmid \expr;\ctx \\
\emph{\textbf{Heaps}} & \heap ::= & \hpemp 
           \spmid \heap\hpjoin\hplocty{\loc}{\evar}{\tyapp{\tycon}{\many{\ty}}}
           \spmid \heap\hpjoin\hplocty{\loc}{\evar}{\tgenobj{\field}{\ty}}
    \\[0.01in]
\emph{\textbf{Function Types}} & \funschema ::= & 
\quant{\many{\loc,\tvar}} \funty{\many{\tybind{\evar}{\ty}}}{\heap}{\equant{\many{\loc'}}\tybind{\evar'}{\ty'}}{\heap'}\\[0.01in]
\end{tabular}
}%
\\
{\small
$\vnum \in \text{Integers}, \ \vref_\loc \in \text{Reference Constants},\ x,y,f \in \text{Identifiers},\ \op \in \{+, -, \ldots\}$
}%
\caption{Syntax of \lang programs and types}
\label{fig:syntax}
\label{fig:types}
\end{figure}
To explain how \toolname infers refinement types as outlined in
\cref{sec:overview}, we first explain the core features of \toolname's
refinement type system.
We focus on the more novel features of our type system; a full
treatment may be found in \iftechrep \cref{sec:annotstatement} \else \cite{arttechrep}\fi.
\subsection{Type Rules}
\ifappendix
\begin{figure}[t!]
\else
\begin{figure}
\fi
\judgementHead{Subtyping}{\issubtype{\ctx}{\heap}{\ty_1}{\ty_2}, {\heapst{\ctx}{\heap}{\heap'}}}
{
$$
\inference
{\validimp{\ctx}{p}{p'}}
{\issubtype{\ctx}{\heap}{\reftp{b}{p}}{\reftp{b}{p'}}}[\rstrule{b}]
$$
\ifappendix
$$
\inference
{
\issubtype{\ctx}{\heap}{\many\ty}{\many\ty'}
\\
{\validimp{\ctx}{p}{p'}}
}
{
\issubtype{\ctx}{\heap}{\reftp{\tyapp{\tycon}{\many\ty}}{p}}
                          {\reftp{\tyapp{\tycon}{\many\ty'}}{p'}}
}[\rstrule{app}]
$$
\[
\inference
{
\issubtype{\ctx}{\heap}{\many\ty}{\many\ty'}
\quad
\validimp{\ctx}{p}{p'}
}
{
\issubtype{\ctx}{\heap}{\reftp{\tgenobj{\field}{\ty}}{p}}
                         {\reftp{\tgenobj{\field}{\ty'}}{p'}}
}[\rstrule{rec}]
\]
\else
{}
\fi
\[
  \inference
    {
      \validimp{\ctx}{p}{p' \wedge \vv \neq \tnull}
    }
    {\issubtype{\ctx}{\heap}{\reftp{\trefm{\loc}}{p}}{\reftp{\tref{\loc}}{p'}}}[\rstrule{down}]
\]

\[
  \inference
    {\validimp{\ctx}{p}{p'}}
    {\issubtype{\ctx}{\heap}{\reftp{\tref{\loc}}{p}}{\reftp{\trefm{\loc}}{p'}}}[\rstrule{up1}]
\;
  \inference
    {\validimp{\ctx}{p}{p'}}
    {\issubtype{\ctx}{\heap}{\reftp{\tnull}{p}}{\reftp{\trefm{\loc}}{p'}}}[\rstrule{up2}]
\]
$$
\inference{
}{
  \heapst{\ctx}
         {\hpemp}
         {\hpemp}
}[\rstheapnull]
\;
\inference{
  \heapst{\ctx}{\heap}{\heap'} \quad
  \heapst{\ctx}{\ty}{\ty'}
}{
  \heapst{\ctx}
         {\heap\hpjoin\hplocty{\loc}{\evar}{\ty}}
         {\heap'\hpjoin\hplocty{\loc}{\evar}{\ty'}}
}[\rstheap]
$$
}

\judgementHead{Heap Folding}{\heapfold{\ctx}{\ty_1}{\heap_1}{\ty_2}{\heap_2}{x}}
{
\[
\inference{
  \locs{\ty_1} \cap \dom{\heap_1} = \emptyset \quad 
  \issubtype{\ctx}{\heap}{\ty_1}{\ty_2}
}{
  \heapfold{\ctx}
             {\ty_1}
             {\heap_1}
             {\ty_2}
             {\heap_2}
             {x}
}[\rstfoldty{base}]
\]
\[
\inference{
  \heap_1 = \heap_1'\hpjoin\hplocty{\loc}{x}{\ty}\quad
  \heap_2 = \heap_2'\hpjoin\hplocty{\loc}{x}{\ty'} \\
  \issubtype{\ctx}{\heap}{\reftp{\tref{\loc}}{p}}{\ty_2} \quad
  \heapfold{\ctx}{\ty}{\heap_1'}{\ty'}{\heap_2'}{x} \\
}{
  \heapfold{\ctx}
           {\reftp{\tref{\loc}}{p}}
           {\heap_1}
           {\ty_2}
           {\heap_2}
           {y}
}[\rstfoldty{ref}]
\]
\[
\inference
{
  \issubtype{\ctx}{\heap}{\reftp{\trefm{\loc}}{p}}{\ty_2} \\
  \heap_1 = \heap_1'\hpjoin\hplocty{\loc}{y}{\ty}\quad
  \heap_2 = \heap_2'\hpjoin\hplocty{\loc}{y}{\ty'} \\
  \heapfold{\tybind{x}
           {\reftp{\trefm{\loc}}{p \wedge \vv \neq null}}
           ;\ctx}
           {\ty}
           {\heap_1'}
           {\ty'}
           {\heap_2'}
           {y}
  \\
  \heapfold{\tybind{x}
           {\reftp{\trefm{\loc}}{p \wedge \vv = null}}
           ;\ctx}
           {\ty}
           {\heap_1'}
           {\ty'}
           {\heap_2'}
           {y}
}
{
  \heapfold{\ctx}{\reftp{\trefm{\loc}}{p}}{\heap_1}{\ty_2}{\heap_2}{x}
}
[\rstfoldty{?ref}]
\]
\[
\inference{
  \heapfold{\ctx}{\ty_i}{\heap_1}{\ty_i'}{\heap_2}{x}
}{
  \heapfold{\ctx}
           {\tgenobj{f_i}{\ty_i}}
             {\heap_1}
             {\tgenobj{f_i}{\ty_i'}}
             {\heap_2}{y}
}[\rstfoldhp]
\]
}
\caption{\ifappendix{Subtyping}\else{Selected subtyping}\fi{}, heap subtyping, and heap folding rules}
\label{fig:heapfolding}
\label{fig:subtyping}
\label{fig:heapsubtyping} 
\end{figure}
\ifappendix
\begin{figure*}
\else
\begin{figure*}[t!]
\fi
\judgementHead{Statement Typing}{\stmttype{\ctx}{\heap}{\stmt}{\ctx'}{\heap'}}
{
\ifappendix
\[
\inference
{
  \exprtype{\ctx}{\heap}{\expr}{\reft}
}
{
  \stmttype{\ctx}{\heap}
           {\evar = \expr}
           {\tybind{\evar}{\reftp{\tau}{\vv = \expr}};\ctx}
           {\heap}
}[\rtassgn]
\]
\ifappendix
\[
\inference
{
\stmttype{\ctx}{\heap}{\stmt_1}{\ctx'}{\heap'}
%
\quad
\stmttype{\ctx'}{\heap'}{\stmt_2}{\ctx''}{\heap''}
}
{
  \stmttype{\ctx}{\heap}{\stmt_1;\stmt_2}{\ctx''}{\heap''}
}[\rtseq]
\else
\inference
{
\stmttype{\ctx}{\heap}{\stmt_1}{\ctx'}{\heap'}
%
\quad
\stmttype{\ctx'}{\heap'}{\stmt_2}{\ctx''}{\heap''}
}
{
  \stmttype{\ctx}{\heap}{\stmt_1;\stmt_2}{\ctx''}{\heap''}
}[\rtseq]
\fi
\]
%
\[
\inference{
  \iswellformed{\ctx}{\heap\hpjoin\hplocty{\loc}{x}{\ty}}
  \quad
  \fresh{\loc,x}
}{
\ifappendix
  \stmttype{\ctx}{\heap}{\pad{\loc}{x}}{\ctx}{\heap\hpjoin\hplocty{\loc}{x}{\ty}}
  \else
  \stmttype{\ctx}{\heap}{\padnone{\loc}}{\ctx}{\heap\hpjoin\hplocty{\loc}{x}{\ty}}
\fi
}[\rtpad]
\]
\else
\fi
\ifappendix
\[
\inference
{
  \exprtype{\ctx}{\heap}{\expr}{\tbool}
  \quad
  \stmttype{\grd{\expr};\ctx}{\heap}{\stmt_1}{\ctx_1;\ctx}{\heap_1}
  \quad
  \stmttype{\grd{\lnot\expr};\ctx}{\heap}{\stmt_2}{\ctx_2;\ctx}{\heap_2}
  \\
  \iswellformed{\ctx}{\heap'}
  \quad
  \ifappendix
  \issubtype{\ctx_1;\ctx}{}{\heap_1}{\heap'}
  \else
  \issubtype{\ctx_1;\ctx}{}{\heap_1}{\heapsubst{\heap'}{\heap_1}}
  \fi
  \quad
  \ifappendix
  \issubtype{\ctx_2;\ctx}{}{\heap_2}{\heap'}
  \else
  \issubtype{\ctx_2;\ctx}{}{\heap_2}{\heapsubst{\heap'}{\heap_2}}
  \fi
 \\
  \mbox{for each $x_i \in \ctx_1 \cap \ctx_2$,}
    \quad
    \exprtype{\ctx_1;\ctx}{\heap_1}{x_i}{\ty_i}
    \quad
    \exprtype{\ctx_2;\ctx}{\heap_2}{x_i}{\ty_i}
  \\
  \iswellformed{\many{\tb{x_i}{\ty_i}};\ctx,\heap'}{\ty_i}
  \quad
  \iswellformed{\many{\tb{x_i}{\ty_i}};\ctx}{\heap'}
  \ifappendix
  { }
  \else
  \quad
  \fresh{\binder{\heap'}}
  \fi
}
{
  \stmttype{\ctx}{\heap}{\ite{\expr}{\stmt_1}{\stmt_2}}{\many{\tb{x_i}{\ty_i}};\ctx}{\heap'}
}[\rtite]
\]
\else
{}
\fi
\[
\inference{
\hastype{\ctx}{\evar}{\tref{\loc}} \quad
        {\hplocty{\loc}{z}{\tgenobj{\field_i}{\ty_i}} \in \heap}
}{
\stmttype{\ctx}
         {\heap}
         {y = \access{\evar}{\field_i}}
         {\tybind{y}{\ty_i};\ctx}
         {\heap}
}[\rtaccess]
\]
\[
\inference{
\hastype{\ctx}{\evar}{\tref{\loc}}
\quad
\hastype{\ctx}{\expr}{\reftp{\tbase}{\pred}}
\\
{\ty_r =  \strfields{z,\tobj{\tybind{\field_{0}}{\ty_{0}}, \ldots,
      \tybind{\field_i}{\reftp{\tbase}{\vv = \expr}}, \ldots}}}
\quad
\fresh{z}
}
{
  \stmttype{\ctx}
{\hplocty{\loc}{y}{\tgenobj{\field_j}{\ty_j}}\hpjoin\heap}
  {\access{\evar}{\field_i} \ifappendix =_z \else = \fi \expr}
  {\ctx}
  {\hplocty{\loc}{z}{\ty_r}\hpjoin\heap}
}[\rtmutate]
\]

\[
\inference{
  \text{for each $\expr_f$, }
    \hastype{\ctx,\heap}{\expr_f}{\ty_f}
  \quad
  \ty = \strfields{z,\tgenobj{\field}{\ty}}
  \quad
  \fresh{\loc,z}
}{
 \stmttype
  {\ctx}
  {\heap}
  {\evar \ifappendix =_z \allocl{\loc}{\many{\field:\expr_f}} 
         \else       =    \alloc{\many{\field:\expr_f}} \fi }
  {\tybind{\evar}{\tref{\loc}};\ctx}
  {\hplocty{\loc}{z}{\ty}\hpjoin\heap}
}[\rtalloc]
\]

\[
\inference{
  \hastype{\ctx,\heap}{x}{\tref{\loc}} \quad
  \ty_y = \reft \quad \ty_{z} = \reftp{\tbase}{\vv = y}
  \quad
  \fresh{z}
}{
  \stmttype
   {\ctx}
   {\hplocty{\loc}{y}{\ty_y} \hpjoin \heap}
   {
   \ifappendix
   {\concBind{x}{z}}
   \else
   {\conc{x}}
   \fi
   }
   {{\tybind{y}{\ty_y}};\ctx}
   {\hplocty{\loc}{z}{\ty_{z}} \hpjoin \heap}
}[\rtconc]
\]
\quad
\[
\inference{
  \typehasdef{\ctx}{\tyapp{\tycon}{\many{\tvar}}}
                   {\tybody{\heap_c}{x_c}{\ty_c}}
  \quad
  \iswellformedMeas{{\tyapp{\tycon}{\many\tvar}}}{\measDef{m}{x}{e_{\fn{m}}}}
  \\
  \heap = \hplocty{\loc}{x}{\reftp{\tyapp{\tycon}{\many{\ty}}}{q}}\hpjoin\heap_0
  \quad
  {\heap' = 
    \hplocty{\loc}{x_c}{ \many{\sub{\tvar}{\ty}}\ty_c}
    \hpjoin
    \many{\sub{\tvar}{\ty}}\heap_c
    \hpjoin
    \heap_0}
  \\
  \iswellformed{\ctx,\heap}{\many{\ty}}
  \quad
  \iswellformed{\ctx,\heap'}{\heap'}
  \quad
  \fresh{\dom{\heap_c},\,\binder{\heap_c},\, {x_c}}
}{
  \stmttype{\ctx}
           {\heap}
           {\ifappendix
           {\unwindBind{\loc}{x_c\cdot\dom{\heap_c}\cdot\binder{\heap_c}}}
           \else
           {\unwind{\loc}}
           \fi
           }
           {(\bigwedge\limits_{\fn{m}} \fn{m}(x) = e_{\fn{m}})
       ;\ctx}{\heap'}
}
[\rtunwind]
\]
\[
\inference{
  \typehasdef{\ctx}{\tyapp{\tycon}{\many{\tvar}}}
                     {\tybody{\heap_c}{\evar}{\ty_c}}
  \\
  \heapfold{\ctx}
           {\ty_x}
           {\heap_x}
           {\many{\sub\tvar\ty}\ty_c}
           {\many{\sub\tvar\ty}\heap_c}
           {x}
  \quad
  \iswellformed{\ctx}{\hplocty{\loc}{y}{\ty_y}\hpjoin\heap'}
  \quad
  \issubtype{\ctx}{}{\heap}{\heap'}
  \\
  \iswellformedMeas{{\tyapp{\tycon}{\many\tvar}}}{\measDef{m}{x}{e_{\fn{m}}}}
  \quad
  \ty_y = {\reftp{\tyapp{\tycon}{\many{\ty}}}{
      \bigwedge_{\fn{m}} \fn{m}(\vv) = e_{\fn{m}}
      }}
  \quad
    \fresh{y}
}{
  \stmttype{\ctx}
           {\hplocty{\loc}{x}{\ty_x}\hpjoin\heap_x\hpjoin\heap}
           {\ifappendix
           {\windBind{y}{\loc}}
           \else
           {\wind{\loc}}
           \fi
           }
           {\ctx}
           {
    \hplocty{\loc}{y}{\ty_y}
    \hpjoin
    \heap'
    }
}[\rtwind]
\]
\ifappendix
\[
\inference{
  \ret{\tybind{\evar_\retId}{\ty_\retId}/\heap_\retId} \in \ctx
  \quad
  \hastype{\ctx}{\expr}{\sub{\evar_\retId}{\expr}\ty_\retId}
  \quad
  \heapst{\ctx}{\heap}{\sub{\evar_\retId}{\expr}\heap_\retId}
}{
  \retstmttype{\ctx}{\heap\hpjoin\heap'}{\return\expr}
}[\rtreturn]
\]

\[
\inference{
  \pgmtype{\Phi}{\fun}{\funschema}
  \quad
  \iswellformed{\ctx,\heap}{\funty{\many{\tybind{x_j}{\ty_j}}}{\heap_i}{\tybind{x_o}{\ty_o}}{\heap_o}
  = \finst{\funschema}{\many{\loc}}{\many{\ty}}}
  \\
  \theta = \many{\sub{x}{e_j}}
  \quad
  (\many{\tybind{y}{\ty_y}},\heap_o') = \addHeap{\tybind{x_o}{\ty_o};\ctx,\heap_o}
  \\
  \text{for each $j$, }
     \exprtype{\ctx}{\heap_u\hpjoin\heap_m}{e_j}{\theta\ty_j}
  \quad
  \heapst{\ctx}{\heap_m}{\theta\heap_i}
}{
  \stmttype{\ctx}
           {\heap_u\hpjoin\heap_m}
           {\evar = \funcall{\many{\expr_j}}}
           {\tybind{x}{\theta\ty_o};\ctx}
           {\heap_u\hpjoin\theta\heap_o}
}[\rtfuncall]
\]
\else
\fi
}
\caption{ \ifappendix Annotated \else Selected \fi Statement Typing Rules. \ifappendix In
  $\rtfuncall$, $\funschema$ is $\alpha$-renamable. Note that the 
  well-formedness check requires all \emph{output} binders to be fresh
  in the calling context. \else \fi
  We assume that type definitions (and, hence, measures over these definitions)
  $\typehasdef{\ctx}{\tyapp{\tycon}{\many{\tvar}}}
  {\tybody{\heap}{\evar}{\ty}}$ are $\alpha$-convertible.  }
\ifappendix
\label{fig:annotstatementtyping}
\else
\label{fig:statementtyping}
\fi
\end{figure*}

\mypara{Type Environments.}
We describe \toolname in terms of an imperative language \lang with
record types and with the usual call by value semantics, whose syntax
is given in \cref{fig:syntax}.
A \emph{function environment} is defined as a mapping, $\Phi$, from
functions $\fun$ to function schemas $\funschema$.
A \emph{type environment} ($\ctx$) is a sequence of \emph{type
  bindings} $\tb{\evar}{\ty}$ and \emph{guard expressions} $e$.
A \emph{heap} ($\heap$) is a finite, partial map from locations
($\loc$) to type bindings.
We write $\ctx(x)$ to refer to $\ty$ where $\tb{x}{\ty} \in \ctx$,
and $\heap(\loc)$ to refer to $\tb{x}{\ty}$ where the mapping
$\loc \mapsto \tb{x}{\ty} \in \heap$.

\mypara{Type Judgements.}
The type system of $\toolname$ defines a judgement
$\pgmtype{\Phi}{\fun}{\funschema}$, which says given the environment
$\Phi$, the function $\fun$ behaves according to its pre- and
post-conditions as defined by $\funschema$.
An auxiliary judgement $\stmttype{\ctx}{\heap}{\stmt}{\ctx'}{\heap'}$
says that, given the input environments \ctx and \heap, \stmt produces
the output environments $\ctx'$ and $\heap'$.
We say that a program $\prog$ \emph{typechecks} with respect to $\Phi$
if, for every function $\fun$ defined in $\prog$,
$\pgmtype{\Phi}{\fun}{\Phi(\fun)}$.

\mypara{Well-Formedness.}
We require that types \ty be \emph{well formed} in their
local environments $\ctx$ and heaps $\heap$, written
$\iswellformed{\ctx,\heap}{\ty}$. 
A heap $\heap$ must heap be well formed in its local environment
$\ctx$, written $\iswellformed{\ctx}{\heap}$.
The rules for the judgment
\iftechrep
(\cref{sec:appendixwf})
\else
\cite{arttechrep}
\fi
capture the intuition that a type may only refer to 
binders in its environment.

\mypara{Subtyping.}
We require a notion of subsumption, \eg 
so that the integer $2$ can be typed either as \reftp{\tint}{\vv=2} 
or simply \tint.
The subtyping relation depends on the environment.
For example, \reftp{\tint}{\vv=x} is a subtype of \reftp{\tint}{\vv=2}
if \tb{x}{\reftp{\tint}{\vv=2}} holds as well.
Subtyping is formalized by the judgment 
\mbox{$\issubtype{\ctx}{}{\ty_1}{\ty_2}$},
of which selected rules are shown in \cref{fig:subtyping}.
Subtyping in \lang reduces to the validity of 
logical implications between refinement predicates.
As the refinements are drawn from a decidable logic of Equality, 
Linear Arithmetic, and Uninterpreted Functions, validity
can be automatically checked by SMT solvers \cite{simplifyj}.
The last two rules convert between non-null 
and possibly null references (\tref{\loc} and \trefm{\loc}).

\mypara{Heap Subtyping.} 
The heap subtyping judgment $\heapst{\ctx}{\heap}{\heap'}$ describes
when one heap is subsumed by another.
\Cref{fig:heapsubtyping} summarizes the rules for heap subsumption.
Heap subtyping is \emph{covariant},
which is sound because our type system is flow sensitive --
types in the heap are updated \emph{after} executing a 
statement.

\mypara{Statements.}
When the condition $x,y$ $\mathit{fresh}$ appears in the antecedent of a rule,
it means that $x$ and $y$ are distinct names that do not appear
in the input environment $\ctx$ or heap $\heap$.
We write $\sub{x}{y}$ for the capture avoiding substitution that maps
$x$ to $y$.
The rules for sequencing, assignment, control-flow joins, and function
calls are relatively straightforward extensions from previous work
(\eg \cite{LiquidPOPL10}).
The complete set of rules may be found in 
\iftechrep
\cref{sec:annotstatement}
\else
\cite{arttechrep}
\fi.

\mypara{Allocation.}
In \rtalloc, a record is constructed from a sequence of 
field name and expression bindings.
The rule types each expression $e_\field$ as $\ty_\field$,
generates a record type $\ty$, and allocates a fresh 
location $\loc$ on the heap whose type is $\ty$. 
To connect fields with their containing records, we 
create a new binder $y$ denoting the record, and use 
the helper $\tyfn{NameFields}$%
\iftechrep 
(\cref{sec:annotstatement})
\else
\cite{arttechrep}
\fi
to \emph{strengthen} the type of each field-binding 
for $y$ from 
$\tybind{\field}{\reftpv{\vv_{\field}}{\tbase}{\pred}}$,
to
$\tybind{\field}{\reftpv{\vv_{\field}}{\tbase}{\pred\wedge\vv_{\field} = \fn{Field}(\vv,\field)}}$. Here, \fn{Field} is an uninterpreted function.

\mypara{Access.}
\rtaccess and \rtmutate both require that \nnull pointers are used
to access a field in a record stored on the heap.
As \rtalloc strengthens each type with \tyfn{NameFields}, the type 
for $y$ in \rtaccess contains the predicate 
$\vv_{f_i} = \fn{Field}(\vv, f_i)$.
Any facts established for $y$ are linked, in the refinement logic,
with the original record's field:
when a record field is \emph{mutated}, 
a \emph{new} type binding is created in the heap, and
each unmutated field is 
linked to the old record using $\fn{Field}$.

\mypara{Concretization.}
As heaps also contain bindings of names to types, it would be
tempting to add these bindings to the \emph{local} environment to
strengthen the subtyping context.
However, due to the presence of possibly null references, 
adding these bindings would be unsound. Consider the program fragment:
\begin{src}
  function f(){ return null; }  
  function g(){ var p = f(); assert(false) }
\end{src}
One possible type for @f@ is 
$
\funty{}{\hpemp}{\equant{\loc}\tb{r}{\trefm{\loc}}}{\hplocty{\loc}{\evar}{\reftp{\tint}{\tfalse}}}
$
because the location $\loc$ is unreachable.
If we added the binding $\tybind{\evar}{\reftp{\tint}{\tfalse}}$ 
to $\ctx$ after the call to \code{f}, then the 
@assert(false)@ in \code{g} would unsoundly typecheck!

We thus require that in order to include a heap binder in a local
context, $\ctx$, the location must first be made \emph{concrete}, 
by checking that a reference to it is definitely \emph{not} null.
Concretization of a location $\loc$ is achieved with the 
\emph{heap annotation} $\concName(x)$.
Given a non-null reference, \rtconc transforms the 
local context $\ctx$ and the heap $\heap$ by
(1)~adding the binding $\tybind{y}{\ty_y}$ at the location $\loc$ to
  $\ctx$;
(2)~adding a \emph{fresh} binding $\tybind{z}{\ty_{z}}$ at $\loc$ that
 expresses the equality $y = z$.

\mypara{Unfold.}
\rtunwind describes how a type constructor application
$\tyapp{\tycon}{\tvar}$ may be unfolded according to its definition.
The context is modified to contain the new heap locations corresponding
to those mentioned in the type's definition.
The rule assumes an $\alpha$-renaming such that the locations and binders
appearing in the definition of $\tycon$ are \emph{fresh}, and then instantiates
the formal type variables $\many\tvar$ with the actual $\many\ty$.
The environment is strengthened using the thus-instantiated measure
bodies.

\mypara{Fold.}
Folding a set of heap bindings \emph{into} a
data structure is performed by \rtwind.
Intuitively, to fold a heap into a type application of \tycon, we
ensure that it is consistent with the definition of \tycon.
Note that the rules assume an appropriate $\alpha$-renaming of the
definition of \tycon.
Simply requiring that the heap-to-be-folded be a
subtype of the definition's heap is too restrictive.  Consider the
first @fold@ in @absL@ in \cref{fig:ex3}.
As we have reached the end of the list \isnull{xn} we need to fold
$$    
      \hb{\x}{\tb{x_1}{\tobj{\fld{data}{\tnat},\fld{next}{\trefm{\addr{t}}}}}} 
\hsep \hb{t}{\tb{t_0}{\tlist{\tint}}}
$$
into $\hb{\x}{\tb{x_2}{\tlist{\tnat}}}$. 
An application of heap subtyping, 
\ie requiring that the heap-to-be-folded is 
a subtype of the body of the type definition,
would require that
$\hb{t}{\tlist{\tint}} \relst \hb{t}{\tlist{\tnat}}$,
%
which does not hold!
However, the fold \emph{is} safe, as  
the @next@ field is \vnull, rendering \addr{t} unreachable.
We observe that it is safe to fold a heap into another heap, so long
as the sub-heap of the former that is \emph{reachable from a given
  type} is subsumed by the latter heap.
 
Our intuition is formalized by the relation
$\heapfold{\ctx,\heap}{\ty_1}{\heap_1}{\ty_2}{\heap_2}{x}$, 
which is read: ``given a local context $\ctx, \heap$, 
the type $\ty_1$ and the heap $\heap_1$ may be folded
into the type $\ty_2$ and heap $\heap_2$.''
$\rstfoldty{base}$ defines the ordinary case:
from the point of view of a type $\ty$, any heap $\heap_1$ 
may be folded into another heap $\heap_2$.
On the other hand, if $\ty_1$ is a reference to a location
$\loc$, then $\rstfoldty{ref}$ additionally requires the 
folding relation to hold at the type bound at $\loc$ in $\heap_1$.

$\rstfoldty{?ref}$ splits into two cases, depending on whether
the reference is null or not.
The relation is checked in two strengthened 
environments, respectively assuming the reference is in fact 
null and non-null.
This strengthening allows the subtyping judgement to make 
use reachability. Recall the first fold in @absL@ that 
happens when \isnull{\pvar{xn}}. To check the @fold(&x)@, the 
rule requires that the problematic heap subtyping 
$\ctx \vdash \hb{t}{\tlist{\tint}} \relst \hb{t}{\tlist{\tnat}}$
only holds when $\mathtt{x.next}$ is non-null, \ie when $\ctx$ is
$$\tb{\pvar{xn}}{\reftp{\trefm{\addr{t}}}{\vv = \pfield{x_2}{next}}},\ 
  \isnull{\pvar{xn}},\ 
  \isnonnull{\pfield{x_2}{next}}$$
This heap subtyping reduces to checking the validity of the following, 
which holds as the antecedent is inconsistent:
$${\pvar{xn}} = \pfield{x_2}{next} \wedge \isnull{\pvar{xn}} \wedge \isnonnull{\pfield{x_2}{next}}
\Rightarrow 0 \leq \vv.$$

\subsection{Refinement Inference}
In the definition of the type system we assumed that type refinements
were given.
In order to \emph{infer} the refinements, we replace each 
refinement in a program with a unique variable, \kvar{_i}, that
denotes the unknown refinement.
More formally, let $\hat{\Phi}$ denote a function environment as
before except each type appearing in $\hat{\Phi}$ is optionally of the
form $\reftpv{\vv}{\tbase}{\kvar{i}}$, \ie its refinement has been
omitted and replaced with a unique \kvar{} variable.
Given a set of function definitions $\prog$ and a corresponding
environment of \emph{unrefined} function signatures $\hat\Phi$, 
to infer the refinements denoted by each \kvar{} we extract a system
of Horn clause constraints $C$.
The constraints, $C$, are \emph{satisfiable} if there exists a mapping
of $K$ of $\kappa$-variables to refinement formulas such each
implication in $KC$, \ie substituting each $\kvar{i}$ with its image
in $K$, is valid.
We solve the constraints by abstract interpretation in the predicate
abstraction domain generated from user-supplied predicate
templates. For more details, we refer the reader
to~\cite{LiquidPLDI08}.
We thus infer the refinements missing from $\hat{\Phi}$ by finding
such a solution, if it exists.

\mypara{Constraint Generation.} Constraint generation
is carried out by the procedure
\fn{CGen} which takes a function environment ($\Phi$), type
environment ($\ctx$), heap environment ($\heap$), and statement
($\stmt$) as input, and ouputs (1)~a set of Horn constraints over
refinement variables $\kappa$ that appear in $\Phi$, $\ctx$, and
$\heap$; (2)~a new type- and heap-environment which correspond to the
effect (or post-condition) after running $\stmt$ from the input type
and heap environment (pre-condition).
\lstdefinestyle{mycaml}{
   language=[Objective]Caml,
   keywordstyle={\bfseries\codesize\sffamily},
   basicstyle=\codesize\sffamily,
   deletekeywords={null},
   mathescape=true,
   columns=flexible,
   keepspaces=true,
   extendedchars=true,
   showstringspaces=false,
   showspaces=false,
   tabsize=2,
   breaklines=true,
   showtabs=false,
   captionpos=b,
   belowskip=0pt
}
\def\inline{\lstinline[style=mycaml,basicstyle=\normalsize\sffamily]}
\lstset{language=Caml}
\lstset{style=mycaml}

\begin{figure}[tb]
\begin{lstlisting}
CGen : FunEnv $\times$ TypeEnv $\times$ HeapEnv $\times$ Stmt $\rightarrow$ {Constr} $\times$ TypeEnv $\times$ HeapEnv
CGen($\Phi$,$\ctx$,$\heap$,s) = match s with
$\ldots$
 | y = x.f $\rightarrow$ let $\ell$ = loc($\ctx$(x)) in ({$\issubtype{\ctx}{}{\ctx(\mathsf{x})}\tref{\loc}$}, y:TypeAt($\heap$,$\ell$);$\ctx$,$\heap$)
 | x.f = e $\rightarrow$ let (cs, t)     = CGEx($\ctx$,$\heap$,e)
                  $\loc$          = Loc(t)
                  (y:T$_{\fn{y}}$, z) = ($\heap$($\loc$), FreshId())
                  ht        = NameFields(z, T$_{\fn{y}}$[f : Shape(t) $\sqcap$ (v = e)])
              in  (cs $\cup$ {$\issubtype{\ctx}{}{\mathsf{t}}{\tref{\loc}}$}, $\ctx$, $\heap$[$\loc \mapsto$z:ht])
\end{lstlisting}
\caption{Statement constraint generation}
\label{fig:cgen1}
\end{figure}

The constraints output by \fn{CGen} correspond to the well-formedness
constraints, $\iswellformed{\ctx, \heap}{\ty}$, and 
subtyping constraints, $\issubtype{\ctx}{}{\ty}{\ty'}$, defined by the
type system.
\emph{Base} subtyping constraints
$\issubtype{\ctx}{}{\reftp{b}{p}}{\reftp{b}{q}}$ correspond
to the (Horn) Constraint $\denote{\ctx} \Rightarrow p \Rightarrow q$,
where $\denote{\ctx}$ is the conjunction of all of the refinements
appearing in $\ctx$ \cite{LiquidPLDI08}.
\emph{Heap} Subtyping constraints $\issubtype{\ctx}{}{\Sigma}{\Sigma'}$
are decomposed via classical subtyping rules into base subtyping
constraints between the types stored at the corresponding locations
in $\Sigma$ and $\Sigma'$. This step crucially allows the predicate
abstraction to sidestep reasoning about reachability and the heap,
enabling inference.

\code{CGen} proceeds by pattern matching on the statement to be typed.
Each @FreshType()@ or @Fresh()@ call generates a new $\kappa$ variable
which may then appear in subtyping constraints as described previously.
Thus, in a nutshell, @CGen@ creates @Fresh@ templates for unknown refinements,
and then performs a type-based \emph{symbolic execution} to generate constraints
over the templates, which are solved to infer precise refinements summarizing
functions and linked structures.
As an example, the cases of \fn{CGen} corresponding to \rtaccess and \rtmutate are
show in \cref{fig:cgen1}.
\subsection{Soundness}
The constraints output by \fn{CGen} enjoy the
following property.
Let ($C$,$\ctx'$,$\heap'$) be the output of
\inline{CGen($\hat{\Phi}$,$\ctx$,$\heap$,s)}.
If $C$ is satisfiable, then there exists some solution $K$ such that
$\stmttypef{K\hat{\Phi}}{K\ctx}{K\heap}{\stmt}{K\ctx'}{K\heap'}$~\cite{LiquidPLDI08},
that is, there is a type derivation using the refinements from $K$.
Thus $K$ yields the inferred program typing $\Phi \myeq K\hat{\Phi}$,
where each unknown refinement has been replaced with its solution,
such that \pgmtype{\Phi}{\fun}{\Phi(\fun)} for each $\fun$ defined in
the program $\prog$.

To prove the soundness of the type system, we translate types,
environments and heaps into separation logic \emph{assertions} and
hence, typing derivations into \emph{proofs} by using the
interpretation function $\denote{\cdot}$. We prove~\iftechrep
(\S~\cref{sec:proof})
\else
\cite{arttechrep}
\fi
the following: 
\showstmtprocthm
$\Pre{\funschema}$, $\Post{\funschema}$ and $\Body{f}$ are the translations
of the input and output types of the function, the function (body) statement.
As a corollary of this theorem, our main soundness result follows:
\showtypeproofs
If we typecheck a program in the empty environment, we get a valid
separation logic proof of the program starting with the pre-condition \ttrue.
We can encode programmer-specified \code{assert}s as calls to a special
function whose type encodes the assertion.
Thus, the soundness result says that if a program typechecks then on
\emph{all} executions of the program, starting from \emph{any} input
state:
\begin{inparaenum}[(1)]
  \item all memory accesses occur on \nnull pointers, and
  \item all assertions succeed.
\end{inparaenum}
%



\lstset{language=JavaScript}
\lstset{style=myjs}
\section{Experiments} \label{sec:eval}
\begin{table*}[tb]\centering
\ra{0.8}
{\small 
\begin{tabular}{l>{\sf}l>{\raggedright}p{5cm}rrr}
\toprule
\textbf{Data Structure} & \textbf{\textrm{Properties}} & \textbf{Procedures}  &
\textbf{LOC} &  \textbf{T}\\
\midrule
  Singly linked list   & Len, Keys & @append@, @copy@, @del@, 
                                     @find@, @insBack@, @insFront@, 
                                     @rev@ & 73 &  2 \\
  Doubly linked list   & Len, Keys & @append@,  @del@, @delMid@,
                                     @insBack@, 
                                     @insMid@, @insFront@ & 90 & 16 \\
  Cyclic linked list   & Len, Keys & @delBack@, @delFront@, @insBack@, @insFront@& 49 & 2 \\
  Sorted linked list   & Len, Keys, Sort & @rev@, @double@, @pairwiseSum@, @insSort@, @mergeSort@, @quickSort@ & 135 & 10 \\
  Binary Tree          & Order, Keys  &  @preOrder@, @postOrder@, @inOrder@ &31 &  2  \\
  Max heap             & Heap, Keys   &  @heapify@ & 48 & 27 \\
  Binary search tree   & BST, Keys    &  @ins@, @find@, @del@ & 105 & 11\\
  Red-black tree       & Red-black, BST, Keys &  @ins@, @del@ 
  & 322 & 213 \\
  \bottomrule
\end{tabular}}%
\caption{Experimental Results (Expressiveness)}
\label{tab:eval}
\end{table*}

\begin{table*}[tb]\centering
\ra{0.9}
{\small
\begin{tabular}{llcccc}
\toprule
\multirow{2}{*}{\textbf{Data Structure}} & \multirow{2}{*}{\textbf{Procedure}}  & \multicolumn{2}{c}{\textbf{ART}} & \multicolumn{2}{c}{\textbf{VCDryad}}\\
& & Specification & Annotation & Specification & Annotation\\
\midrule
Singly Linked List & (definition) & 34  & - & 31  & -\\
&   @rev@       & 5    & 0      & 11 & 15
\\
Sorted Linked List & (definition) & 38 & - & 50 & -\\
& @rev@               & 11   & 9      & 17 & 15
\\
& @double@       & 0    & 4  & 7  & 54
\\
& @pairwiseSum@ & 0    & 4  & 13 & 75
\\
& @insSort@           & 5    & 0      & 20 & 17
\\ 
& @mergeSort@        & 5    & 18 & 18 & 79
\\ 
& @quickSort@        & 5    & 18 & 11 & 140
\\
Binary Search Tree & (definition) & 58 & - & 55 & - \\
& @del@              & 7    & 32     & 20 & 33
\\
\midrule
\multicolumn{2}{r}{\textbf{Total}} & 168 & 63 & 253 & 428 \\
\bottomrule
\end{tabular}
\caption{Experimental results (Inference). For each procedure listed we compare the number of tokens used to specify:
  \textbf{ART} Type refinements for the top-level procedure in \toolname; 
  \textbf{ART} Annot manually-provided predicate templates required to infer the necessary types \cite{LiquidPLDI08};
  \textbf{VCDryad} Spec pre- and post-conditions of the corresponding top-level VCDryad procedure; and
  \textbf{VCDryad} Annot loop invariants as well as the specifications required for intermediate functions in VCDryad. \toolname Annot totals include only \emph{unique} predicate templates across benchmarks.
\label{tab:infereval}
}
}%
\end{table*}

We have implemented alias refinement types in a tool called \toolname.
The user provides (unrefined) function signatures, and \toolname
infers
(1)~annotations required for alias typing, and
(2)~refinements that capture correctness invariants.
We evaluate \toolname on two dimensions:
the first demonstrates 
that it is \emph{expressive} enough to verify a variety of
sophisticated properties for linked structures;
the second that it provides a significant \emph{automation} over the
state-of-the-art, represented by the SMT-based \vcd system.
\vcd has annotations comparable to other recent tools that use
specialized decision procedures to discharge Separation Logic
VCs~\cite{jafarPLDI15}.
Our benchmarks are available at \cite{artbenchs}.

\mypara{Expressiveness.} Table~\ref{tab:eval} summarizes the set of
data structures, procedures, and properties we used to evaluate the
expressiveness of \toolname.
The user provides the type definitions, functions (with
unrefined type signatures), and refined type specifications
to be verified for top-level functions, \eg the top-level
specification for @insertSort@.
\textbf{LOC} is lines of code and \textbf{T}, the verification time in seconds.

We verified the following properties, where applicable:
%
  [{\fn{Len}}] the output data structures have the expected length;
  [{\fn{Keys}}] the elements, or ``keys'' stored in each data structure
  [{\fn{Sort}}] the elements are in sorted order
  [{\fn{Order}}] the ouput elements have been labeled in the correct order
  (\eg preorder)
  [{\fn{Heap}}] the elements satisfy the max heap property
  [{\fn{BST}}] the structure satisfies the binary search tree property
  [{\fn{Red-black}}] the structure satisfies the red-black tree property.

\mypara{Automation.}
To demonstrate the effectiveness of \emph{inference}, we
selected benchmarks from \cref{tab:eval} that made use of
loops and intermediate functions requiring extra proof
annotations in the form of pre- and post-conditions in
\vcd, and then used type inference to infer the
intermediate pre- and post-conditions.
The results of these experiments is shown in \cref{tab:infereval}.
We omit incomparable benchmarks, and those where the
implementations consist of a single top-level function.
We compare the number of tokens required to specify type
refinements (in the case of \toolname) and pre- and
post-conditions (for \vcd).
The table distinguishes between two types of annotations: (1)~those
required to specify the desired behavior of the top-level procedure,
and (2)~additional annotations required (such as intermediate function
specifications).
Our results suggest that it is possible to verify the correctness of a
variety of data-structure manipulating algorithms without requiring
many annotations beyond the top-level specification.
On the benchmarks we examined, overall annotations required by
\toolname were about 34\% of those required by \vcd.
Focusing on intermediate function specification, \toolname required
about 21\% of the annotation required by \vcd.

\mypara{Limitations.}
Intuitively, \toolname is limited to ``tree-like''
ownership structures: while sharing and cycles
are allowed (as in double- or cyclic-lists),
there is a tree-like \emph{backbone} used
for traversal.
For example, even with a singly linked list,
our system will reject programs that traverse
deep into the list, and return a pointer to a
cell \emph{unboundedly} deep inside the list.
We believe it is possible to exploit the connection
made between the SL notion of ``magic wands" and
the type-theoretic notion of ``zippers" \cite{Huet97}
identified in \cite{summersECOOP15} to enrich the
alias typing discipline to accommodate such access patterns.



\section{Related Work}


\mypara{Physical Type Systems.}
\toolname infers logical invariants in part by leveraging the
technique of alias typing~\cite{AliasTypesRec,LinLocFI07},
in which access to dynamically-allocated memory is factored
into references and capabilities.
In \cite{chargueraud08,mezzo}, capabilities are used to
decouple references from regions, which are collections of values.
In these systems, algebraic data types with an ML-like ``match''
are used to discover spatial properties, rather than \tnull pointer
tests.
\windName \& \unwindName are directly related to
roll \& unroll in \cite{AliasTypesRec}.
These operations, which give the program access to
quantified heap locations, resemble reasoning about
capabilities \cite{plaid,mezzo}.
These systems are primarily restricted to verifying
(non-)aliasing properties and finite, non-relational
facts about heap cells (\ie ``typestates"), instead of
functional correctness invariants.
A possible avenue of future work would be to use a more sophisticated
physical type system to express more data structures with sharing.

\mypara{Logical Type Systems.}
Refinement types~\cite{XiPfenning99,nystromSPG2008,LiquidPLDI09},
encode invariants about recursive algebraic data types using indices
or refinements.
These approaches are limited to \emph{purely functional} languages,
and hence cannot verify properties of linked, mutable structures.
\toolname brings logical types to the imperative setting
by using \cite{AliasTypesRec} to structure and reason about
the interaction with the heap.

\mypara{Interactive Program Logics.}
Several groups have built interactive verifiers and used them to
verify data structure correctness~\cite{kuncak08,CohenTPHOL09}.
These verifiers require the programmer write pre- and postconditions
and loop invariants in addition to top-level correctness
specifications.
The system generates verification conditions (VCs)
which are proved with user interaction.
\cite{jacobs2011verifast} uses symbolic execution and SMT solvers
together with user-supplied tactics and annotations to prove programs.
\cite{HTT,Chlipala11} describe separation logic frameworks for Coq
and tactics that provide some automation.
These are more expressive than \toolname but
require non-trivial user assistance to prove VCs.

\mypara{Automatic Separation Logics.}
To automate the proofs of VCs (\ie entailment),
one can design decision procedures for various
fragments of SL, typically restricted to common
structures like linked lists.
\cite{berdine2006smallfoot} describes an entailment
procedure for linked lists, and
\cite{predator,haase2013seloger,BouajjaniSLAD2012}
extend the logic to include constraints on list data.
\cite{LahiriQadeer08,botinvcanPS2009,navarro2011separation,piskac2013automating}
describe SMT-based entailment by reducing formulas
(from a list-based fragment) to first-order logic,
combining reasoning about shape with other SMT theories.
The above approaches are not extensible (\ie limited to list-segments);
other verifiers support user defined, separation-logic predicates, with
various heuristics for entailment~\cite{ChinDNQ12,jafarPLDI15}.
\toolname is related to natural proofs~\cite{Qiu0SM13,VCDryad} and
the work of Heule et al. ~\cite{heuleverification},
which instantiate recursive predicates using the
local footprint of the heap accessed by a procedure,
similar to how we insert \windName and \unwindName
heap annotations, enabling generalization and
instantiation of structure properties.
Finally, heap binders make it possible to use recursive
functions (e.g. measures) over ADTs in the imperative
setting. While our measure instantiation~\cite{LiquidPLDI09}
requires the programmer adhere to a typing discipline,
it does not require us to separately prove that the
function enjoys special properties~\cite{KuncakPOPL10}.

\mypara{Inference.}
The above do not deal with the problem of
inferring annotations like the inductive invariants
(or pre- and post- conditions) needed to generate
appropriately strong VCs.
To address this problem, there are several abstract
interpreters~\cite{TVLA} tailored to particular
data structures like list-segments~\cite{cookCAV08},
lists-with-lengths~\cite{MagillTHOR08}.
Another approach is to combine separate domains for heap and data
with widening strategies tailored to particular structures~\cite{gulwaniPOPL08,changPOPL08}.
These approaches conflate reasoning
about the heap and data using monolithic assertions or abstract domains,
sacrificing either automation or expressiveness.

\mypara{\ackname}
This work was supported by NSF grants CCF-1422471, C1223850, CCF-1218344, and 
a generous gift from Microsoft Research.

{
\bibliographystyle{plain}
\bibliography{main}
}
\include{main.bib}

\iftechrep
\appendixtrue
\appendix
%
\FloatBarrier
\section{\lang with Annotations}\label{sec:annotlang}
\subsection{Annotation Elaboration}
In \cref{sec:soundness} we state that our soundness proof relies
on a semantics-preserving elaboration step that inserts assignments
to ghost variables.
More formally, we can take a typing derivation for a program in \lang
as presented in \cref{sec:cgen} and produce an elaborated source with
an almost identical derivation.
These elaborations amount to making assignments to \emph{ghost}
variables to correspond to \emph{locations} and \emph{heap binders},
and are thus semantics preserving because the ghost variables do not
appear in the original program.
The type judgements for the elaborated code are used in the soundness
proof.
The syntax of \lang with ghost variable annotations is given in
\cref{fig:annotsyntax}.  
Statement typing judgements with ghost variable annotations are given
in \cref{fig:annotstatementtyping}.
 
\begin{figure}
{\small
  \ra{0.9}
  \begin{tabular}{l >{$}l<{$} >{$}l<{$}}
\emph{\textbf{Expressions}} & \expr ::= &
                   \vnum 
           \spmid  \vtrue \spmid \vfalse
           \spmid \vnull
           \spmid  \vref_\loc
           \spmid  \evar
           \spmid  \expr \op \expr \\[0.01in]
\emph{\textbf{Statements}} & \stmt ::= &
                       \stmt; \stmt 
            \spmid      \evar = \expr
            \spmid      y = \access{\evar}{\field}
            \spmid      \access{\evar}{\field} \ifappendix =_z \else = \fi \expr
            \spmid      \ite{\expr}{\stmt}{\stmt}
            \\
           &&
            \spmid      \return{\expr}
           \spmid     \evar \ifappendix =_z \allocl{\loc}{\many{\field:\expr}} 
                              \else = \alloc{\many{\field:\expr}} \fi 
            \spmid      \evar = \funcall{\many{\expr}}
            \\
           &&
            \spmid      \ifappendix \unwindBind{\loc}{\many{y}} \else \unwind{\loc} \fi
            \spmid      \ifappendix \windBind{\loc}{x} \else \wind{\loc} \fi
            \spmid      \ifappendix \concBind{x}{y} \else \conc{x} \fi
          \ifappendix
           \spmid     \pad{\loc}{x}
          \else
           \spmid     \padnone{\loc}
          \fi
\\
\emph{\textbf{Programs}} & \prog ::= & \many{\fundecl{f}{\many{\evar}}}\\
\end{tabular}
}%
 
{\small
\ra{0.9}
\begin{tabular}{l>{\ } >{$}l<{$} >{$}l<{$} >{$}l<{$}}
\emph{\textbf{Primitive Types}} & \tprim ::= & 
                    \tint
            \spmid  \tbool
            \spmid  \tvar
            \spmid  \tnull
            \spmid  \tref{\loc}
            \spmid  \trefm{\loc}
            \\[0.01in]
\emph{\textbf{Types}} & \tbase ::= &
                  \tprim
           \spmid \tyapp{\tycon}{\many{\ty}}
           \spmid \tgenobj{\field}{\ty}\\[0.01in]
\emph{\textbf{Refined Types}} & \ty ::= & \reft\\[0.01in]
\emph{\textbf{Type Definition}} & \tycon ::=    & 
                      \tydef{\tyapp{\tycon}{\many{\tvar}}}
                            {\heap}
                            {\evar}
                            {\tobj{\many{f:\ty}}}\\[0.01in]
\emph{\textbf{Contexts}} & \ctx ::= & \emptyset \spmid \tybind{\evar}{\ty};\ctx 
                                \spmid \expr;\ctx \\
\emph{\textbf{Heaps}} & \heap ::= & \hpemp 
           \spmid \heap\hpjoin\hplocty{\loc}{\evar}{\tyapp{\tycon}{\many{\ty}}}
           \spmid \heap\hpjoin\hplocty{\loc}{\evar}{\tgenobj{\field}{\ty}}
    \\[0.01in]
\emph{\textbf{Function Types}} & \funschema ::= & 
\quant{\many{\loc,\tvar}} \funty{\many{\tybind{\evar}{\ty}}}{\heap}{\equant{\many{\loc'}}\tybind{\evar'}{\ty'}}{\heap'}\\[0.01in]
\end{tabular}
}%

\caption{Syntax of \lang programs and types}
\label{fig:annotsyntax}
\label{fig:annottypes}
\end{figure}

\label{sec:annotstatement}
\appendixtrue

\appendixfalse

\section{Well Formedness and Expression Typing}
\appendixtrue
\label{sec:appendixwf}
\ifappendix
\begin{figure}
\else
\fi
\judgementHead{Well Formed Types}{\iswellformed{\ctx,\heap}{\ty}}
\[
\inference{
  \fv{\pred} \subseteq 
    \fv{\tybind{\vv}{\tprim};\ctx}\cup \fv{\heap}\quad
  \pred \mbox{ is well-sorted in $\tybind{\vv}{\tprim};\ctx$ and
    $\heap$}
}{
  \iswellformed{\ctx,\heap}{\reftp\tprim\pred}
}[\wf{prim}]
\]
\[
\inference
{
  \tydef{\tyapp{\tycon}{\many{\tvar}}}
                         {\heap}
                         {\evar}
                         {\tobj{\many{\tybind{f}{\ty}}}}\quad
  \iswellformed{\ctx,\heap}{q}\\
  \fn{length}(\many{\tvar}) = \fn{length}({\reftp{\tprim_i}{\pred_i}})\quad
  \iswellformed{\ctx,\heap}{\many{\reftp{\tprim_i}{\pred_i}}}\quad
}
{
  \iswellformed{\ctx,\heap}{\reftp{\tyapp{\tycon}{\many{\reftp{\tprim_i}{\pred_i}}}}{q}}
}[\wf{app}]
\]
\[
\inference
{
  \iswellformed{\ctx,\heap}{\many{\reftp{\tprim_i}{\pred_i}}}\quad
  \iswellformed{\ctx,\heap}{q}\quad
  \mbox{$\field_i = \field_{i'} \Rightarrow i = i'$}
}
{
  \iswellformed{\ctx,\heap}
               {\reftp{\tgenobj{\field_i}{\reftp{\tprim_i}{\pred_i}}}{q}}
}[\wf{rec}]
\]
\caption{Well Formed Types}
\end{figure}
\begin{figure}
\judgementHead{Well Formed Type Definitions}{\iswellformed{}{\tycon}}
\[
\inference{
  \iswellformed{\heap}{\ty}\quad
  \iswellformed{\tybind{\evar}{\reftp{\tobj{\many{\tybind{f}{\ty}}}}{p}},\heap}{\heap}\\
  \evar \notin \fv{\heap}\quad
  \fn{TypeVars}(\heap) \cup \fn{TypeVars}(\reftp{\tobj{\many{\tybind{f}{\ty}}}}{p}) \subseteq \many{\tvar}\quad
}{
  \iswellformed{}{\tydef{\tyapp{\tycon}{\many{\tvar}}}
                        {\heap}
                        {\evar}
                        {\reftp{\tobj{\many{\tybind{f}{\ty}}}}{p}}}
}[\wf{def}]
\]
\caption{Well Formed Type Definitions}
\end{figure}
\begin{figure}
\judgementHead{Well Formed Heaps and Worlds}
 {\iswellformed{\ctx}{\heap}\quad\iswellformed{\ctx,\heap}{\heap'}
  \quad\iswellformed{\ctx,\heap}{\wld{\evar}{\ty}{\heap'}}}
\[
\inference{
}{
  \iswellformed{\ctx,\heap}{\hpemp}
}[\wf{emp}]
\quad
\inference{
\iswellformed{\ctx,\heap}{\ty} \quad
\iswellformed{\ctx,\heap}{\heap'}
\\
\loc \notin \dom{\heap'}
\quad   
\evar \notin \binder{\heap'} \cup \binder{\ctx}
}{
\iswellformed
  {\ctx,\heap}
  {\hplocty{\loc}{\evar}{\ty}\hpjoin\heap'}
}[\wf{bind}]
\]
\[
\inference{
  \iswellformed{\ctx,\heap}{\heap}
}{
  \iswellformed{\ctx}{\heap}
}[\wf{heap}]
\]
\[
\inference{
\evar \notin \ctx
\quad
\iswellformed{\tybind{\evar}{\ty};\ctx,\heap}{\ty}
\\
\iswellformed{{\tybind{\evar}{\ty}};\ctx,\heap}{\heap'}
}{
\iswellformed{\ctx}
             {\wld{\evar}{\ty}{\heap'}}
}[\wf{wld}]
\]
\caption{Well Formed Heaps and Worlds}
\end{figure}
\begin{figure}
\judgementHead{Well Formed Schemas}{\iswellformed{\ctx,\heap}{\funschema}}
\[
\inference{
\text{for each $j$, $x_j \notin \fv{\ctx} \cup \fv{\heap}$}
\text{ and } 
  \iswellformed{\many{\tybind{\evar_j}{\ty_j}};\ctx,\heap_i}{\ty_j}
\\
\iswellformed{\many{\tybind{\evar_j}{\ty_j}};\ctx,\heap}{\heap_i}
\quad
\iswellformed{\many{\tybind{\evar_j}{\ty_j}};\ctx}
             {{\tybind{\evar_o}{\ty_o}}/\heap_o}
\quad
x_o \notin \fv{\ctx} \cup \fv{\heap}
\\
\fv{\heap_o} \nsubseteq \fv{\ctx} \cup \fv{\heap}
}{
\iswellformed
{\ctx,\heap}
{\quant{\many{\loc,\tvar}}
\funty{\many{\tybind{\evar_j}{\ty_j}}}{\heap_i}{\equant{\many{\loc_o}}\tybind{\evar_o}{\ty_o}}{\heap_o}}
}[\wf{fun}]
\]
\caption{Well Formed Schemas}\label{fig:wellformed}
\end{figure}
\begin{figure}
\input{measurejudge}
\caption{Well Formed Measures}
\end{figure}

\label{sec:appendixexpr}
\ifappendix
\begin{figure}
\else
\begin{figure}[ht!]
\fi
  \judgementHead{Expression Typing}{\exprtype{\ctx}{\heap}{\expr}{\ty}}
\[
\inference{}{\exprtype{\ctx}{\heap}{\vnum}{\reftp{\tint}{\vv = \vnum}}}[\rtint]
\quad
\inference{
  \tybind{\evar}{\reftp{\tbase}{\pred}} \in \ctx
}
{
  \exprtype{\ctx}{\heap}{\evar}{\reftp{\tbase}{\vv = \evar}}
}[\rtvar]
\]
\[
\inference{}{\exprtype{\ctx}{\heap}{\vnull}
{\reftp{\tnull}{\vv = \vnull}}}[\rtnull]
\]
\[
\inference
{\exprtype{\ctx}{\heap}{\expr_1}{\tint}\quad\exprtype{\ctx}{\heap}{\expr_2}{\tint}}
{\exprtype{\ctx}{\heap}{\expr_1\op\expr_2}{\reftp{\tint}{\vv = \expr_1 \op \expr_2}}}
[\rtop]
\]
\[
\inference
{\exprtype{\ctx}{\heap}{\expr}
          {\reftp{\tref{\loc}}{p}}}
{\exprtype{\ctx}{\heap}{\expr}
          {\reftp{\tref{\loc}}{p \wedge \vv \neq \vnull}}}[\rtrefinv]
\]
\[
\inference{\exprtype{\ctx}{\heap}{\expr}{\ty_1}
           \quad\issubtype{\ctx}{\heap}{\ty_1}{\ty_2}
           \quad\iswellformed{\ctx,\heap}{\ty_2}}
          {\exprtype{\ctx}{\heap}{\expr}{\ty_2}}[\rtsub]
\]
\caption{Expression Typing Rules}\label{fig:exprtyping}
\end{figure}

\begin{figure}
\judgementHead{Function Typing}{\pgmtype{\Phi}{\fun}{\funschema}}
$$
\inference{
  \many{\loc} = dom(\heap_i) \quad \many{\loc_o} = dom(\heap_o) \setminus dom(\heap_i) \\
  \iswellformed{\emptyset,\hpemp}{\funschema = \quant{\many{\loc}}\quant{\many{\tvar}}\funty{\many{\tybind{\evar}{\ty}}}{\heap_i}{\equant{\many{\loc_o}}\tybind{\evar_o}{\ty_o}}{\heap_o}} \\
  (\ctx, \heap) = \addHeap{\many{\tybind{\evar}{\ty}},\heap_i}\\
  \retstmttype{\ret{\tybind{\evar_o}{\ty_o}/\heap_o};\ctx}
           {\heap}
           {\stmt}
}{
  \pgmtype{\Phi}{\fun}{\funschema}
}[\rtfun]
$$
\caption{Function Typing Rules}
\label{fig:functiontyping}
\end{figure}

\newcommand{\heaptheta}{
\begin{cases}
  x' & \hplocty{\loc}{x}{\ty} \in \heap \mbox{\textit{ and }}
       \hplocty{\loc}{x'}{\ty'} \in \heap'\\
  x & \mbox{\textit{otherwise}}
\end{cases}
}
\begin{figure*}
\emph{\bf Substitution of Heap Binders}
\begin{flalign*}
\heapsubst{\heap}{\heap'} & = \theta\heap'
\quad \mbox{\textit{where}} \quad
{\theta}x = \heaptheta&
\end{flalign*}
%
%
%
\\
\textbf{Fresh Renaming of Worlds}
\begin{flalign*}
&
\begin{array}{ll}
  \freshsub{\wld{x}{\ty}{\heap}} & = \fn{FreshSubst'}(x,\ty,\heap,\hpemp)\\
\fn{FreshSubst'}(x,\ty,\hpemp,\theta) &= \sub{x}{x'} \cdot \theta \quad \fresh{x}\\
\fn{FreshSubst'}(x,\ty,\hplocty{\loc}{y}{\ty_y}\hpjoin\heap,\heap') & = 
  \fn{FreshSubst'}(x,\ty,\heap,\sub{\loc}{\loc'}\cdot\sub{y}{y'}\cdot\theta) \quad \fresh{\loc',y'}
\end{array}&
\end{flalign*}
%
\\
\textbf{Strengthening Field Names}
\begin{flalign*}
&\strfields{\reftpv{\vv}{\tgenobj{\field}{\reftpv{\vv_f}{\tbase_f}{\pred_f}}}{\pred}} =
\reftpv{\vv}{\tgenobj{\field}{\reftpv{\vv_f}{\tbase_f}{\pred_f \wedge \vv_f = \objfield{\vv}{\field}}}}{\pred}&
\end{flalign*}
\\
\textbf{Function Instantiation}
\begin{flalign*}
&\begin{array}{ll}
\finst{\quant{\loc}\funschema}{\loc'\cdot\many{L}}{\many{\ty}} & = 
  \finst{\sub{\loc}{\loc'}\funschema}{L}{\many{\ty}} \\
\finst{\quant{\tvar}\funschema}{\epsilon}{\ty\cdot\many{\ty}} & = 
  \finst{\sub{\tvar}{\ty}\funschema}{\epsilon}{\many{\ty}} \\
\finst{\funschema}{\epsilon}{\epsilon} = \funschema
  %
\end{array}&
\end{flalign*}
\\
\textbf{Local Context Strenghtening}
\begin{flalign*}
\addHeap{\ctx,\heap} & = 
\begin{cases}
(\ctx,\heap) & \heap = \hpemp\\
(\tybind{y}{\ty};\ctx', \hplocty{\loc}{z}{\reftp{\tbase}{\vv = y}}\hpjoin\heap')
  & \heap = \hplocty{\loc}{y}{\reft}\hpjoin\heap_0,\\
  & \tybind{x}{\tref{\loc}} \in \ctx,\ 
  \mbox{\textit{and $z$ fresh}}\\
(\ctx',\heap') & \mathit{otherwise}
\end{cases}&\\
  & \mathit{where} \quad (\ctx', \heap') = \addHeap{\ctx,\heap_0}&
\end{flalign*}
\caption{Helper Functions}
\label{fig:helper}
\end{figure*}

Well-formedness rules are given in this section.
Expression typing judgements are given in \cref{fig:exprtyping}.

\FloatBarrier
\setcounter{lemma}{0}
\setcounter{theorem}{0}
\setcounter{corollary}{0}
\section{Soundness of \textsc{ART}}
\label{sec:proof}

\section{Soundness}\label{sec:soundness}

We translate 
types, environments and heaps into separation logic \emph{assertions}
and hence, typing derivations into separation logic \emph{proofs}.
We illustrate the translation by example and present the 
key theorems; see~\cref{sec:proof} for details.

\subsection{Translation}
We define a translation $\denote{\cdot}$ 
from types, contexts, and heaps to assertions 
in separation logic. 
We translate a typing context $\ctx,\heap$ by conjoining the 
translations of $\ctx$ and $\heap$.
Next, we illustrate the translation of local contexts $\ctx$ 
and heaps $\heap$ using the \code{insert} function from \cref{fig:ex4}.
When we describe \eg a $\ctx$ at a program point, we are referring
to the $\ctx$ \emph{before} that program point's execution.

\mypara{Local Contexts}
We translate contexts $\ctx$ into a conjunction of 
\emph{pure}, \ie heap independent, assertions.
The context $\ctx_0$ \emph{before} \pgmpoint{A} contains two
assumptions: (1)~\x is a possibly \vnull
reference to \addr{\x} and (2)~\x is definitely \vnull.  Note
that \addr{\x} \emph{does not} denote the application of a function
called \addr{}; it is a distinct location \emph{named} \addr{\x}.  We
translate $\ctx_0$ by conjoining the translations of these contents.
Note that translation states that if \x is 
not \vnull, then \x refers to \addr\x.
\begin{align*}
\ctx_0 \doteq \tb{\x}{?\tref{\addr{\x}}},\ \isnull{\x} &&
\denote{\ctx_0} \doteq (\x \neq \vnull \Rightarrow \x = \addr{\x}) \wedge (\x = \vnull)
\end{align*}

\mypara{Heaps}
We interpret heaps as (separated) conjunctions of \emph{impure} 
(\ie, heap-dependent assertions) and \emph{pure} assertions 
corresponding to the types of the values stored in each location.
The heap \emph{before} \pgmpoint{A}
states that if the location exists on the heap, 
 then it refers a \emph{valid list}, described 
by $\seplista{A}{\addr\x}{x_o}$.
\begin{align}
  \heap_0         \doteq  \hb{\x}{\tb{x_0}{\tlist{A}}} &&
\denote{\heap_0}  \doteq  (\addr{\x} \neq \vnull) \Rightarrow \seplista{A}{\addr{\x}}{x_0}
 \label{eq:trans:heap1}
\end{align}

\mypara{Type Predicates}
We translate the definition of the $\tlist{A}$ type into 
a separation logic assertion that characterizes the
part of the heap where a $\tlist{A}$ is stored.
For example, for the list type from \cref{fig:ex4} we define 
the type predicate as:
\begin{align*}
  \seplista{A}{\loc}{x}  \doteq
      & \ \sepunfold{\loc}{x} \wedge
      \exquant{h, t, \loc_t} \headlista{A}{\loc}{h}{\loc_t} \hpjoin 
      ((\loc_t \neq \vnull) \Rightarrow \seplista{A}{\loc_t}{t}) \\
  \headlista{A}{\loc}{h}{\loc_t} \doteq 
      & \  \denote{\hplocty{\loc}{h}{\tobj{\tybind{\fn{data}}{A}
                                          ,\tybind{\fn{next}}{\trefm{\loc_t}}}}
                  } 
\end{align*}
Intuitively, the conjuncts of the assertion $\seplista{A}{\loc}{x}$ say:
\begin{inparaenum}[(1)]
  \item the snapshot $x$ is a \emph{pure value} comprising all the 
        records (data and references) that are transitively reachable 
        on the heap starting at $\loc$, and
  \item the heap satisfies the ``shape" invariants defined by the recursive
        alias type \tlist{A}.
\end{inparaenum}
%
%
We prove that each well-formed type definition has
a translation analagous to the above, allowing
us translate heap-locations mapped to recursive types in a manner
analagous to~\cref{eq:trans:heap1}.

\mypara{Records} 
The other case is for heap-locations that are mapped to plain records.
At \pgmpoint{B}, \code{insert} has 
allocated a record to which \y refers.
There, we have\\
\vspace{0.25\baselineskip}
\begin{minipage}{0.25\linewidth}
\raggedleft
\begin{equation}
\ctx_1 \doteq \ctx_0, \tb{\y}{\addr{\y}} \label{eq:ctx2}
\end{equation}
\end{minipage}
\begin{minipage}{0.75\linewidth}
\raggedright
\begin{equation}
\denote{\ctx_1} \doteq \denote{\ctx_0} \wedge (y \neq \vnull \Rightarrow y = \addr{y}) \wedge (y \neq \vnull)  \label{eq:ctx2:trans}
\end{equation}
\end{minipage}
\\
\\
\ie the binding for \y is translated to the assertion for a
maybe-null pointer plus the fact that \y is not-null. The heap before
executing
\pgmpoint{B} is translated:
\begin{equation}
\denote{\heap_1} \doteq \denote{\heap_0} \hpjoin (\sepmap{\addr{\y}}{y_0} \wedge {y_0 = \tobj{\fld{data}{\code{k}},\fld{next}{\vnull}}})
\label{eq:heap2}
\end{equation}

\subsection{Soundness Theorem}
We prove soundness by showing that a valid typing of a statement \stmt
under given contexts can be translated into a valid separation logic
proof of \stmt under pre- and post-conditions obtained by translating
the contexts.
Note that there an (semantics-preserving) elaboration step that, for
clarity, we defer to \cref{sec:proof} which inserts
\emph{ghost variables} to name locations and heap binders.

\showstmtprocthm

$\Pre{\funschema}$, $\Post{\funschema}$ and $\Body{f}$ are the translations
of the input and output types of the function, the function (body) statement.
By virtue of how contexts are translated, the result of the translation is 
\emph{not} a vacuous separation logic proof. 
As a corollary of this theorem, our main soundness result follows:
\showtypeproofs
If we typecheck a program in the empty environment, we get a valid 
separation logic proof of the program starting with the pre-condition \ttrue.
We can encode programmer-specified \code{assert}s as calls to a special 
function whose type encodes the assertion. 
Thus, the soundness result says that if a program typechecks then on
\emph{all} executions of the program, starting from \emph{any} input
state:
\begin{inparaenum}[(1)]
  \item all memory accesses occur on \nnull pointers, and
  \item all assertions succeed.
\end{inparaenum}

The soundness proof proceeds by induction on the typing derivation. 
We show that the typing rules correspond to separation logic proofs
using the appropriate proof rule.
The critical parts of the proof are lemmas that translte the rules for 
subtyping and folding recursive type definitions into
separation logic entailments:

\showfolding
\setcounter{lemma}{\value{lemma}-1}
\refstepcounter{lemma}
\label{ll:folding}
\showsubtyping
\setcounter{lemma}{\value{lemma}-1}
\refstepcounter{lemma}
\label{ll:subtyping}
These lemmas effectively translate 
\windName, \unwindName, and subtyping (\cref{fig:subtyping})
into a combination of ``skip'' statements and the consequence rule
of separation logic.

For example, at \pgmpoint{B}, \code{insert} is about to 
return the list that \y points to (at location \addr{\y}). 
At this point the context is given by the $\ctx_1$ 
and $\heap_1$ from \cref{eq:ctx2,eq:heap2}. 
\toolname checks that the context satisfies the output type of 
\code{insert} \ie that \addr{\y} is the root of a valid list whose 
length equals one greater than (the snapshot of) the input \x.
It uses subtyping and \rtwind to obtain at \pgmpoint{C}: 
$$\ctx_2   \doteq \ \ctx_1, \mlen{y_1} = 1 + \mlen{\vnull} \ \  
  \heap_2  \doteq \ \heap_0 \hpjoin \hb{\y}{\tb{\y_1}{\tlist{A}}}$$
Using \cref{ll:folding} and \cref{ll:subtyping} we have:
${\denotef{\ctx_1}{\heap_1} \Rightarrow \denotef{\ctx_2}{\heap_2}}$
In particular, note that $\y_1$ names the snapshot value
corresponding to the (single-celled) list referred to by 
\y, and the measure instantiation due to \rtwind states 
that the length of the snapshot is 1 greater than the 
length of the tail \vnull.
The implication follows from the definitions
of the length measure, the snapshot predicate and the type 
predicate. 
We then use the implication in an instance of the consequence rule.

Soundness of \toolname follows from the following:
\begin{inparaenum}[(1)]
  \item
    data type definitions correspond to an inductively defined set of 
    \emph{pure} snapshot values;
  \item
    heap binders correspond to snapshot values;
  \item
    no data structure is modified without first \emph{unfolding} it,
    ensuring that each ``snapshot'' (or ``version'') receives a fresh name.
\end{inparaenum}

\subsection{Assertion language}

Values in \lang are records, integers (which are also used as addresses),
the constant \vnull, or products of integers and records. 
We assume an intuitionistic interpretation of assertions (and thus
$p \hpjoin true \Leftrightarrow p$). 
\begin{align*}
  {\values} & = \mathbb{Z} \cup \mathbb{O} \cup \{ null \}&\\
  \mathbb{O} & = \mathsf{Records}&\\
  \mathbb{Z} & = \mbox{\textsf{Integers (and addresses)}}&\\
  \mathbb{S} & = \mathbb{Z} \times \mathbb{O}
\end{align*}
\begin{tabular}{l>{$}l<{$} >{$}c<{$} >{$}l<{$} l}
\textbf{\emph{Expressions}}&  E & ::= & \cdots & \\
&   & \spmid & \tgenobj{\field}{E} & \emph{record constant} \\
&   & \spmid & \mathsf{f(\many{E})} & \emph{UIF application} \\
    \\
\textbf{\emph{Assertions}}&   P & ::=  & \cdots &  \\
&      & \spmid  & \hpemp & \emph{empty heap} \\
&      & \spmid  & \sepmap{E}{E} & \emph{singleton heap} \\
&      & \spmid  & P \hpjoin P & \emph{separating conjunction}
  \end{tabular}\\

With the following axiom:
\begin{equation}
(y = \tobj{\ldots \tybind{\field}{e} \ldots})
\Rightarrow
(P \Leftrightarrow \sub{e}{\objfield{y}{\field}}P)
\label{a:field}
\end{equation}
       
\subsection{\lang Proof Rules}

In addition to the standard frame rule and consequence rule, we assume 
the following axioms:

\judgementHead{Proof Rules}{\htriplefun{F}{P}{s}{Q}}

\mypara{Concretization}
\[
\inference{
  z \notin \fv{e}\quad x,z\ \mathit{distinct}
}{
  \htriple{
    \sepmap{x}{e}
  }{
    \concBind{x}{z}
  }{
    \sepmap{x}{e} \wedge z = e
  }
}
\]

\mypara{Unfolding}
\[
\inference{
}{
  \htriple{
    \exquant{\many{x}}P
  }{\unwindBind{\loc}{\many{x}}}{
    P
  }
}
\]

\mypara{Folding}
\[
\inference{
 x \mbox{ does not appear free in $P$ }
}{
  \htriple{\exquant{x}P}{\windBind{\loc}{x}}{P}
}
\]

\mypara{Padding}
\[
\inference{
  \loc,x\ \mathit{distinct}
}{
  \htriple{\hpemp}{\pad{\loc}{x}}{\hpemp \wedge \loc = \vnull \wedge x = \vnull}
}
\]

\mypara{Assignment}
\[
\inference{
  x \notin \fv{e}
}{
  \htriple{\hpemp}{x = e}{x = e}
}
\]

\mypara{Conditional}
\[
\inference{
  \htriplefun{F}{B \wedge P}{s_1}{Q} \quad \htriplefun{F}{\neg B \wedge P}{s_2}{Q}
}{
  \htriplefun{F}{P}{\ite{B}{s_1}{s_2}}{Q}
}
\]

\mypara{Allocation}
\[
\inference{
  x,z \notin \many{\fv{\expr}}
}{
  \htriple
  {
    \hpemp
  }
  {\evar =_z \alloc{\many{\field:\expr}}}
  {
    \evar = \loc
    \wedge \loc \neq \vnull
    \wedge
    \sepmap{\evar}{z} \wedge z = {\tgenobj{\field}{\expr}}
  }
}
\]

\mypara{Access}
\[
\inference
{
  a,v\ \mathit{distinct\ vars}
}
{ 
  \htriple
  {
   \sepmap{a}{\tgenobj{\field}{\expr}}
  }
  {v = \access{a}{\field_i}}
  {
   \sepmap{a}{\tgenobj{\field}{\expr}}
   \wedge  
   v = \expr_i
  }
}
\]

\mypara{Mutation}
\[
\inference
{
  z \notin \fv{v}
}
{ 
  \htriple
  {
   \sepmap{a}{\tobj{\ldots\tybind{\field_i}{-}\ldots}}
  }
  {\access{a}{\field_i} =_z v}
  {
    \sepmap{a}{z} \wedge z = 
{\tobj{\ldots\tybind{\field_i}{v}\ldots}}
  }
}
\]

\mypara{Sequence}
\[
\inference{
  \htriplefun{F}{P}{s}{R} \quad \htriplefun{F}{R}{s'}{Q}
}{
  \htriplefun{F}{P}{s;s'}{Q}
}
\]


\mypara{Procedure Call}
\[
\inference
{
  g(x_1 \ldots x_m) \doteq s; \return{x_o}\\
  \htriplefun{\hproofrule{P}{g(x_1 \ldots x_m)}{Q},F}{P}{s; \return{x_o}}{Q}
}
{
  \htriplefun{F}{P}{x = g(x_1 \ldots x_m)}{x = x_o \wedge Q}\\
}
\]

\mypara{Return}
\[
\inference
{
}
{
  \htriple{\sub{x_o}{e}P}{\return{e}}{P}
}
\]


\subsection{Definitions}\label{sec:defs}
\begin{definition}[Base type translation]
\begin{flalign*}
&\begin{array}{ll}
\sbase{\tint}{\evar} & = \suif{int}(\evar) \\ 
\sbase{\tnull}{\evar} & = (\evar = \vnull) \\
\sbase{\tref{\loc}}{\evar} & = ((\evar = \loc) \wedge (\evar \neq \vnull))\\
\sbase{\trefm{\loc}}{\evar} & = 
  (\sbase{\tref\loc}{\evar} \vee \sbase{\tnull}{\evar}\\
\sbase{\tobj{\many{\tybind{\field_i}{\ty_i}}}}{\evar} & =
  (\evar = \tgenobj{\field_i}{\objfield{\evar}{\field_i}}\\
\end{array}&
\end{flalign*}
\end{definition}

\begin{definition}[Local type binding translation]
\begin{flalign*}
&\begin{array}{ll}
  \sprop{\reftt{\tprim}}{x} & = \sub{\vv}{x}\pred \wedge \sbase{\tprim}{x}\\
  \sprop{\reftt{\tyapp{\tycon}{\ty}}}{x} & = \sub{\vv}{x}\pred\\
  \sprop{\reftt{\tobj{\many{\tybind{\field_i}{\ty_i}}}}}{x} & =
  \sub{\vv}{x}\pred \wedge \sbase{\tobj{\many{\tybind{\field_i}{\ty_i}}}}{x} 
  \wedge
  \bigwedge\limits_{\tybind{\field_i}{\ty_i}\in \tbase}
    \!
    \sprop{\ty_i}{\objfield{\evar}{\field_i}}
\end{array}&
\end{flalign*}
\end{definition}

\begin{definition}[Heap type binding translation]
\[
\begin{array}{ll}
  \shprop{\reftt{\tyapp{\tycon}{\ty}}}{\loc}{x} & = 
  ((\loc \neq \vnull) \Rightarrow (\sprop{\reftt{\tyapp{\tycon}{\ty}}}{x}) \hpjoin \shtapp{\ty}{\loc}{\evar})
  \\
  & \wedge\ ((\loc = \vnull) \Rightarrow (x = \vnull))\\
  \shprop{\reftt{\tgenobj{\field_i}{\ty_i}}}{\loc}{x} & = 
  ((\loc \neq \vnull) \Rightarrow
    (\sprop{\reft}{\evar} \hpjoin \sepmap{\loc}{\evar})\\
  & \wedge\ ((\loc = \vnull) \Rightarrow (x = \vnull))
\end{array}
\]
\end{definition}

\begin{definition}[Free Variables of \ctx,\heap]
  \[
  \begin{array}{ll}
\fn{FV}(\emptyset)  & =\emptyset \\
\fn{FV}(\tybind{x}{\trefm{\loc}};\ctx) & = \{ \loc \} \cup \fn{FV}(\ctx)\\
\fn{FV}(\tybind{x}{\tref{\loc}};\ctx) &= \{ \loc \} \cup \fn{FV}(\ctx) \\
& \\
\fn{FV}(\hpemp) &= \emptyset \\
\fn{FV}(\hplocty{\loc}{x}{\ty}\hpjoin\heap) &= \{\loc, x \} \cup \fn{FV}(\heap)
\end{array}
\]
\end{definition}


\begin{definition}[Pure Value Types] In order to show measure well
  formedness, we extend the language of types with products ($\ty
  \times \ty'$) and unions with \tnull ($\ty + \tnull$), and define:
\[
\snaptyN : \fn{Type} \times \fn{Heap} \rightarrow \fn{Type}
\]
\[
\begin{array}{ll}
  \snapty{\tref{\loc}}{\heap} & =
    \begin{cases}
      \tref{\loc} \times \snapty{\ty}{\heap} &
        \hplocty{\loc}{x}{\ty} \in \heap\\
      \tref{\loc} &
        {\loc} \notin \dom{\heap}\\
    \end{cases}\\
  \snapty{\trefm{\loc}}{\heap} & =
    \snapty{\tref{\loc}}{\heap} + \snapty{\tnull}{\heap}\\
  \snapty{\ty}{\heap} & = \ty
\end{array}
\]

\end{definition}

\begin{definition}[Pure Values from Type Definitions]
For each type $\ty$ we denote the pure values associated with that type
as $\purety{\ty}$, which we define inductively:
\[
\inference{}{\tnull \in \purety{\tnull}}
\inference{e \in \mathbb{Z}}{e \in \purety{\tint}}
\inference{e \in \values}{e \in \purety{\tvar}}
\]
\[
\inference{
  e \in \mathbb{Z}
}{
  e \in \purety{\tref{\loc}}
}
\inference{e \in \purety{\tnull}}{e \in \purety{\trefm{\loc}}}
\inference{
  e \in \purety{\tref{\loc}}
}{
  e \in \purety{\trefm{\loc}}
}
\]

\[
\inference{
  e \equiv \tgenobj{\field_i}{\ty_i}
  \quad
  e \in \purety{\ty_i}
}{
  \tgenobj{\field_i}{e_i} \in \purety{\tgenobj{\field_i}{\ty_i}}
}
\inference{
e_1 \in \purety{\ty_1}
\quad
e_2 \in \purety{\ty_2}
}{
  (e_1, e_2) \in \purety{(\ty_1 \times \ty_2)}
}
\inference{
e \in \purety{\ty_1} \vee e \in \purety{\ty_2}
}{
e \in \purety{(\ty_1 + \ty_2)}
}
\]

\[
\inference{
  \tydef{\tyapp{\tycon}{\tvar}}{\heap}{\evar}{\ty}
  \\
  e \in \purety{(\snapty{\sub{\tvar}{\ty_{\tvar}}\ty}{\sub{\tvar}{\ty_{\tvar}}\heap} + \tnull)}
}{ 
  e \in \purety{\tyapp{\tycon}{\ty_{\tvar}}}
}
\]
\end{definition}


\begin{definition}[Constraints from Pure Values]
  We consider elements of $\purety{\tycon}$ as ``snapshots'' of some
  heap. We ``restore'' these snapshots by mapping them to \emph{assertions}
  with the functions 
  \begin{align*}
    \sepunfoldN &: \mathbb{Z} \times \values \rightarrow \fn{Prop}\\
    \walkfnN &: \values \rightarrow \values \times \power{\mathbb{Z} \times \values}
  \end{align*}
\begin{align*}
\sepunfold{\loc}{x} & \myeq 
  ((\loc \neq \vnull) \Rightarrow
  (\sepmap{\loc}{e} \circledast_{(\loc',e') \in h} \sepmap{\loc'}{e'}))\\
  \mbox{where } (e,h) & = \walkfn{x}
\end{align*}
and where \fn{Walk} is defined as follows:
\begin{align*}
\walkfn{\tgenobj{\field}{e}}
& = (\tgenobj{\field}{e'}, \bigcup \many{h})
\quad
\mbox{where} \quad \many{(e',h)} = \many{\walkfn{e}}\\
\walkfn{(e_1, e_2)} & = (e_1, \{ (e_1, e_3) \} \cup h) \quad
\mbox{where} \quad (e_3, h) = \walkfn{e_2}\\
\walkfn{e} & = (e, \emptyset)
\end{align*}
\end{definition}
\begin{definition}[Assertions from Type Definitions]
Given the definition
$\tydef{\tyapp{\tycon}{\tvar}}{\heap}{x}{\ty}$, 
define the assertion
\begin{align*}
  &\shtapp{\ty_{\tvar}}{\loc}{x} \myeq
    (\exquant{x_c,\fv{\heap}}
      (\sepmap{\loc}{x_c} \wedge \sprop{\sub{\tvar}{\ty_{\tvar}}\ty}{x_c})
      \hpjoin
      \denote{\sub{\tvar}{\ty_{\tvar}}\heap} 
    \wedge 
    \sepunfold{\loc}{x} 
\end{align*}
\end{definition}

\begin{definition}[Interpretation of Measures]
  Let $\fn{m}(\tybind{x}{\tyapp{\tycon}{\tvar}}) \myeq e$ and $s$ be a mapping from variables to values. Define
  $\sepmeas{\fn{m}(x)} \myeq \sepmeas{e}$ where
  \begin{flalign*}
    \sepmeas{v}(s)&= v &\\
    \sepmeas{\access{x}{f}}(s) &= s(x).f\\
    \sepmeas{\fn{f}(e)}(s) &= f(\sepmeas{e}(s)) &\\
    \sepmeas{e_1 \oplus e_2}(s) &= \sepmeas{e_1}(s) \oplus \sepmeas{e_2}(s) & \\ &\oplus \in \{ +, -, \ldots \}&\\
    \sepmeas{\ite{e}{e_1}{e_2}}(s) &=
    \begin{cases}
      \sepmeas{e_1}(s) & \sepmeas{e}(s) = true \\
      \sepmeas{e_2}(s) & otherwise 
    \end{cases}&\\
  \end{flalign*}
\end{definition}

\begin{definition}[Interpretation of type contexts/worlds]
  \begin{flalign*}
    \denote{\ctx} & = 
      \bigwedge_{{\expr \in \ctx}} \expr
      \bigwedge_{\mathclap{\tybind{\evar}{\ty} \in \ctx}}  \sprop{\ty}{\evar}
      \bigwedge_{\mathclap{\code{type}\dots \in \ctx}}  \sepdef{\code{type}\;t = T} \\
    \denote{\heap} & =
      \bigast_{\hplocty{\loc}{x}{\ty} \in \heap} 
      \shprop{\ty}{\loc}{x} \\
    \denotef{\ctx}{\heap} & = 
    \denote{\ctx} \wedge \denote{\heap}
  \end{flalign*}
\end{definition}

\begin{definition}[Interpretation of procedure declarations]
Assume that $\fun(x) \myeq s$. 
Given $\funschema =
{\quant{\many{\loc}}\quant{\many{\tvar}}\funty{\many{\tybind{\evar}{\ty}}}{\heap_i}{\equant{\many{\loc_o}}\tybind{\evar_o}{\ty_o}}{\heap_o}}
$,
\begin{flalign*}
  \Pre{\funschema} &=  \denote{\many{\tybind{x}{\ty}}}\wedge\denote{\heap_i}&\\
  \Post{\funschema} &= \denote{{\tybind{x_o}{\ty_o}}}\wedge\denote{\heap_o}&\\
  \Body{\fun} &= s&
\end{flalign*}
and the variables appearing in $\Post{\funschema}$ and not
$\Pre{\funschema}$ are considered to be modified by $\fun$.
\end{definition}

\begin{definition}[Interpretation of procedure contexts]
\begin{flalign*}
\denote{\tybind{f}{\funschema};\Phi} 
  & = \hproofrule{\Pre{\funschema}}{\Body{\fun}}{\Post{\funschema}};\denote{\Phi}&
\end{flalign*}
\end{definition}

\subsection{Type Soundness}
We assume the following:
\begin{enumerate}
\item The set of program variables ($x$,$y$, \etc) is disjoint from
  the set of symbols used to denote locations ($\loc$, $\loc'$, \etc).
\item All programs are in single static assignment form. The only
  variables which have more than one static assignment are ``phi''
  variables which are assigned once in each branch of an ``if''
  statement.
\end{enumerate}
%

\showwfctx
\begin{proof}
  By induction on the type well-formedness judgement.
\end{proof}

\begin{corollary}
  The location $\loc$ only appears in $\denote{\ctx}$ if there exists $\tybind{x}{\ty} \in \ctx$
  and $\ty = \reftp{\tref{\loc}}{p}$ or $\ty = \reftp{\trefm{\loc}}{p}$.
\end{corollary}

\showwfheap
\begin{proof}
  By definition of $\denote{\heap}$ and induction on the well-formedness judgement.
\end{proof}

\showschemawf
\begin{proof}
 By the assumptions of $\wf{fun}$
\end{proof}

\begin{corollary}
  The location $\loc$ only appears in $\denote{\heap}$ if there exists $\hplocty{\loc'}{x}{\ty} \in \heap$
  and $\loc' = \loc$ or $\ty = \reftp{\tref{\loc}}{p}$ or $\ty = \reftp{\trefm{\loc}}{p}$.
\end{corollary}


\showwfmeasures
\begin{proof}
  By induction on the $\hastypeMeas{\ctx}{e}{\ty}$ judgement.
\end{proof}

\showsubtyping
\begin{proof}
By induction on the subtyping derivation.

\begin{enumerate}
\item[\textbf{Case} $\issubtype{\ctx}{}{\reft}{\reftp{\tbase}{p'}}$] \hfill \\
By assumption.

\myline
\item[\textbf{Case} $\issubtype{\ctx}{}{\reftp{\tyapp{\tycon}{\many\ty}}{p_\tycon}}
                                         {\reftp{\tyapp{\tycon}{\many\ty'}}{p_\tycon'}}$] \hfill \\
By assumption,
\[ \denote{\ctx} \wedge p_\tycon \Rightarrow p_\tycon' \]
and
\[ \denote{\ctx} \wedge \sprop{\ty}{\vv} \Rightarrow \sprop{\ty'}{\vv} \]
for each $\ty$, $\ty'$. 
%
By hypothesis and the definition of \spropn, 
and by the form of each \ty,
each $\pred$ only occurs
positively in $\shtapp{\many{\reftp{\tau}{\pred}}}{\loc}{\evar}$.
Destructing
$\many{\ty} \mbox{ as } \many{\reftp{\tau}{\pred}}$ 
and
$\many{\ty'} \mbox{ as } \many{\reftp{\tau}{\pred'}}$:
\begin{align*}
\sprop{\reftp{\tyapp{\tycon}{\many\ty}}{p_\tycon}}{\vv} 
& \myeq
p_\tycon \wedge \exquant{\loc}\shtapp{\many{\ty}}{\loc}{\vv}\\
& \Rightarrow p_\tycon' \wedge \exquant{\loc}{\shtapp{\many{\ty'}}{\loc}{\vv}}\\
& \Leftrightarrow \sprop{\reftp{\tyapp{\tycon}{\many{\ty'}}}{p_\tycon'}}{\vv}
\end{align*}

\myline
\item[\textbf{Case} 
  $\issubtype{\ctx}{\heap}{\reftp{\tgenobj{\field}{\ty}}{p}}
  {\reftp{\tgenobj{\field}{\ty'}}{p'}}$] \hfill \\
   Unfolding the definition of \spropn,
   \begin{align*}
   \sprop{\reftp{\tgenobj{\field}{\ty}}{p}}{\vv} = 
   p &
   \wedge
   \vv = \tgenobj{\field}{\objfield{\field}{\vv}}\\
    &\wedge
   \many{\sprop{\ty}{\objfield{\vv}{\field}}}
   \end{align*}
   By hypothesis,
   \[ \denote{\ctx} \Rightarrow \sprop{\ty}{\vv} \Rightarrow \sprop{\ty'}{\vv} \]
   for each $\ty$, $\ty'$, and applying \cref{a:field} gives us
   \[ \denote{\ctx} \Rightarrow \sprop{\ty}{\objfield{v}{\field}} \Rightarrow \sprop{\ty'}{\objfield{v}{\field}} \]
   Now,
   \begin{align*}
     \sprop{\ty_1}{\vv} & = 
       p \wedge \tgenobj{\field}{\objfield{\vv}{\field}}
         { }\wedge
         \bigwedge\limits_{\tybind{\field}{\ty}} \sprop{\ty}{\objfield{\vv}{\field}}
   \end{align*}
   so, combined with the assumption that $\denote{\ctx} \Rightarrow p \Rightarrow p'$,
   \begin{align*}
     \denote{\ctx} \Rightarrow
       p \wedge 
         \tgenobj{\field}{\objfield{\vv}{\field}}
         \wedge
         \bigwedge\limits_{\tybind{\field}{\ty'}} \sprop{\ty'}{\objfield{\vv}{\field}}
   \end{align*}
   which is equivalent to
   $\reftp{\tgenobj{\field}{\ty'}}{p'}$

\myline
  \item[\textbf{Case}
    $\issubtype{\ctx}{\heap}{\reftp{\trefm{l}}{p}}{\reftp{\tref{l}}{p'}}$] \hfill \\
    By assumption,
    \[ \denote{\ctx} \Rightarrow p \Rightarrow (p' \wedge \vv \neq null) \]
    By definition,
    \[ 
      \sprop{{\reftp{\trefm{l}}{p}}}{\vv} = p \wedge (\vv = l
      \wedge \vv \neq null) \vee (\vv = null) 
    \] 
    and thus, combined with the assumption,
    \[ \denote{\ctx} \Rightarrow p \Rightarrow
      (p' \wedge \vv = l
      \wedge \vv \neq null)
    \] 
    which implies
    \[ \sprop{{\reftp{\tref{l}}{p'}}}{\vv} = p' \wedge \vv = l \wedge \vv \neq null \]

\myline
  \item[\textbf{Case}
  $\issubtype{\ctx}{\heap}{\reftp{\tref{l}}{p}}{\reftp{\trefm{l}}{p'}}$]
  \hfill \\
    By assumption,
    \[ \denote{\ctx} \Rightarrow p \Rightarrow (p' \wedge \vv \neq null \wedge \vv = l) \]
    and thus
    \[ \denote{\ctx} \Rightarrow p \Rightarrow (p' \wedge ((\vv \neq null \wedge \vv = l) \vee \vv = null)) \]
    which is equivalent to \sprop{\reftp{\trefm{l}}{p'}}{\vv}.
    
\myline
  \item[\textbf{Case}
  $\issubtype{\ctx}{\heap}{\reftp{\tnull}{p}}{\reftp{\trefm{l}}{p'}}$]
  \hfill \\
    By assumption,
    \[ \denote{\ctx} \Rightarrow p \Rightarrow (p' \wedge \vv = null) \]
    and thus
    \[ \denote{\ctx} \Rightarrow p \Rightarrow (p' \wedge ((\vv \neq null \wedge \vv = l) \vee \vv = null)) \]
    which is equivalent to \sprop{\reftp{\trefm{l}}{p'}}{\vv}.

\myline
  \item[\textbf{Case} $\issubtype{\ctx}{}{\hpemp}{\hpemp}$] \hfill \\
    The heaps are equivalent and their domains are empty, so the conclusion is 
    trivially true.

\myline
  \item[\textbf{Case} 
    $\issubtype{\ctx}{}{\heap\hpjoin\hplocty{\loc}{\evar}{\ty}}
    {\heap'\hpjoin\hplocty{\loc}{\evar}{\ty'}}$]
    \hfill \\
    By the inductive hypothesis, 
    \[
    \denote{\ctx} \Rightarrow \denote{\heap} \Rightarrow \denote{\heap'}
    \]
    and by assumption and the definition of $\shpropn$,
    \[
    \denote{\ctx} \Rightarrow
    \shprop{\ty}{\loc}{x} \Rightarrow \shprop{\ty'}{\loc}{x}
    \]
    hence,
    \[
    \denote{\ctx} \Rightarrow
\shprop{\ty}{\loc}{x} \hpjoin \denote{\heap} \Rightarrow 
\shprop{\ty'}{\loc}{x}
\denote{\heap'}
    \]
    and thus,
    \[
    \denote{\ctx} \Rightarrow
    \denote{\hplocty{\loc}{x}{\ty} \hpjoin \heap} 
    \Rightarrow 
    \denote{\hplocty{\loc}{x}{\ty'} \hpjoin \heap'} 
    \]
  \end{enumerate}
  \qed
\end{proof}

As a consequence of our interpretation of typing contexts, the
following lemma immediately follows:

\showdecomp
\begin{proof}
  Follows immediately from the definition of $\denote{-}$.
\end{proof}

\showexprtypelemma
\begin{proof}
  By induction on the expression typing judgment. In particular, we
  make use of the fact that the typing and subtyping rules for pure
  expressions only depend on $\ctx$. Subsumption requires application of \cref{l:subtyping}.
\end{proof}

\showptrtype
\begin{proof}
  Follows from \cref{l:wfheap} and the definition of $\denote{-}$.
\end{proof}


\showfolding
%
\begin{proof}
By induction on the judgement derivation.

\begin{enumerate}
\item[\textbf{Case} \rstfoldty{base}] \hfill \\
By the subtyping hypothesis,
\[ \denote{\ctx} \wedge \shprop{\ty_a}{\foldLoc}{x} \Rightarrow \shprop{\ty_2}{\foldLoc}{x} \]
And thus, by letting every other $\loc \in \dom{\heap_2} = null$,
\[ \exquant{\fv{\heap_2}}\denote{\foldExtra{\ty_2}\hpjoin\heap_2}. \]
Since $\ty_1$ contains no pointers, and since we are assuming the
satisfiability of $\sprop{\ty_1}{x}$, we can construct $y$ from the
base type of $\ty_1$ and its refinement.

\myline
\item[\textbf{Case} \rstfoldty{ref}] \hfill \\
By the subtyping hypothesis, where $\ty_1 = \reftp{\tref{\loc'}}{p}$,
\[ 
\denote{\ctx} \Rightarrow \sprop{\ty_1}{x} \Rightarrow \sprop{\ty_2}{x}
\]
%
And by the inductive hypothesis,
\[ 
\denote{\ctx} \Rightarrow 
  \denote{\hplocty{\loc'}{x}{\ty}\hpjoin\heap'_1} \Rightarrow
    \exquant{\fv{\heap'_2}}\denote{\hplocty{\loc'}{x}{\ty'}\hpjoin\heap'_2}
    \wedge
    \exquant{s}{\sepunfold{\loc'}{e}}
\]
Combining the two gives us
\[ \exquant{\fv{\heap_2}}\denote{\hplocty{\loc}{y}{\ty_2}\hpjoin\heap_2}. \]
We construct $y = (\loc', s)$ so that $y \in \purety{\tref{\loc'}}$ to yield
\[
\sepunfold{\loc}{y}
\]
\myline
\item[\textbf{Case} \rstfoldty{?ref}] \hfill \\
By hypothesis,
\begin{align*}
&  ((x \neq null) \wedge \foldHypolhs{\ctx}{\ty}{\heap_1'}) \Rightarrow \\
& \quad (\foldHyporhs{y'}{\ty'}{\heap_2'})\\
&  ((x = null) \wedge \foldHypolhs{\ctx}{\ty}{\heap_1'}) \Rightarrow \\
& \quad (\foldHyporhs{y'}{\ty'}{\heap_2'})
\end{align*}
In the first case, $(x \neq null)$
we let 
\[ y = (\loc, y'). \]
In the second case,
\[ y = null. \]
We combine both hypotheses and $y$ constructions to conclude:
\[ \exquant{\fv{\heap_2}}\denote{\foldExtra{\ty_2}\hpjoin\heap_2}. \]
and, by definition, $\sepunfold{\loc}{y}$,

\myline
\item[\textbf{Case} \rstfoldhp]
By hypothesis,
\begin{align*}
& \foldHypolhs{\ctx}{\ty}{\heap_1}) \Rightarrow \\
& \quad (\foldHyporhs{x_i}{\ty}{\heap_2})
\end{align*}
for each $\objfield{x}{\field_i}$ in $\ty_1$.
If we substitute $\objfield{x}{\field_i}$ for $x$ in each conclusion, we conclude
\[
\denote{\hplocty{\loc}{x}{\ty_2}\hpjoin{\heap_2}}
\]
and we construct $y = \tgenobj{\field_i}{x_i}$.
\end{enumerate}
\qed
\end{proof}

\showpure
\begin{proof}
By contrapositive. Assume $x_f \notin \purety{\tycon(A)}$. Then 
\[ x_f  \notin \purefn{A}{\purety{\tycon}}{\ty}{\heap}. \]
The proof is by induction on the derivation of 
$x_f \notin \purefn{A}{\purety{\tycon}}{\ty}$.
The base case starts with the root of $\sepunfold{\loc}{x_f}$ 
\[ \sepunfold{\loc}{x_f} = \sepmap{\loc}{e} \ast \cdots \]
If $x_v \notin \purefn{A}{\purety{\tycon}}{\ty}{\heap}$ then either
$e$ is not a record value, trivially showing
$\neg\denote{\hplocty{\loc}{\evar}{\ty}\hpjoin{\heap}}$, or $x_v$
contains some field $f$ with value $e_f$ such that.
\begin{enumerate}[(i)]
  \item $\tybind{f}{\ty_f} \in \ty$;
  \item and $e_f \notin \purefn{A}{\purety{\tycon}}{\ty_f}{\heap}$
\end{enumerate}
Assuming $e_f \neq (\loc', e'),$ if $e_f \notin \purefn{A}{\purety{\tycon}}{\ty_f}{\heap}$,
then the assertion in $\sepunfold{\loc}{x_f}$,
$\sepmap{\loc}{\tobj{\ldots \tybind{f}{e_j} \ldots}}$, must not satisfy
the corresponding $\denote{\hplocty{\loc}{x}{\ty}}$ (by the definition
of $\purefn{A}{\purety{\tycon}}{\ty}{\heap}$.
If $e_f = (\loc', e')$ for some $\loc', e'$, then 
$e' \notin \purefn{A}{\purety{\tycon}}{\ty_f}{\heap}$
However, by the same reasoning as above, 
$\sepunfold{\loc'}{e'}$ must not be equisatisfiable with 
$\denote{\hplocty{\loc}{\evar}{\ty}\hpjoin{\heap}}$
\end{proof}

\showstmtthm
\begin{proof} 
  By induction on the typing derivation of $\stmt$.

  \begin{enumerate}
  \item[\textbf{Case} $x = e$:] \hfill \\ 
    By assumption, $x$ does not appear in $\ctx$. 
Since $e$ is well-typed, $x \notin \fv{e}$.
By the frame rule, since $x \notin \ctx$ and $x \notin \heap$,
we derive the following rule:
\begin{align*}
\inference{}{
  \htriple{\denotef{\ctx}{\heap}}
          {x = e}
          {\denotef{\ctx}{\heap}\hpjoin x = e}
}
\end{align*}
We then perform the deduction by the definition of $\sprop{\cdot}{\cdot}$:
\begin{align*}
&\hassert{\denotef{\ctx}{\heap} \wedge (x = e)}\\
&\hassert{\denotef{\ctx}{\heap} \wedge \sprop{\reftp{\tbase}{\vv = e}}{x}}\\
&\hassert{\denotef{\tb{x}{\reftp{\tbase}{\vv = e}};\ctx}{\heap}}
\end{align*}

    \myline
  \item[\textbf{Case} $s;s'$] \hfill \\
    By hypothesis,
\begin{align*}
  \htriplefun{\Phi}{\denotef{\ctx}{\heap}}{s_1}{\denotef{\ctx'}{\heap'}}
\end{align*}
and
\begin{align*}
  \htriplefun{\Phi}{\denotef{\ctx'}{\heap'}}{s_2}{\denotef{\ctx''}{\heap''}}
\end{align*}
By straightforward application of the sequence rule, we immediately conclude
\begin{align*}
  \htriplefun{\Phi}{\denotef{\ctx}{\heap}}{s_1;s_2}{\denotef{\ctx''}{\heap''}}
\end{align*}

    \myline
  \item[\textbf{Case} $\ite{e}{s_1}{s_2}$:] \hfill \\
    By hypothesis:
\begin{align*}
  \htriplefun
  {\Phi}
  {e \wedge \denotef{\ctx}{\heap}}{s_1}{\denotef{\ctx_1;\ctx}{\heap_1\hpjoin\heap_1'}}\\
  \intertext{ and }
  \htriplefun
  {\Phi}
  {\neg e \wedge \denotef{\ctx}{\heap}}{s_2}{\denotef{\ctx_2;\ctx}{\heap_2\hpjoin\heap_2'}}
\end{align*}
For each $x_i$, by \cref{l:subtyping}
\begin{align*}
  \denote{\ctx_1;\ctx} \Rightarrow \sprop{\ty'}{x_i} \quad \mathit{and} \quad
  \denote{\ctx_2;\ctx} \Rightarrow \sprop{\ty'}{x_i}
\end{align*}
and by \cref{l:subtyping}
\begin{align*}
  \denote{\ctx_1;\ctx} \Rightarrow \denote{\heap_1} 
                   \Rightarrow \denote{\heap'}
\quad \mathit{and} \quad
  \denote{\ctx_2;\ctx} \Rightarrow \denote{\heap_2} \Rightarrow \denote{\heap'}
\end{align*}
Combining these facts allows us to conclude, %
\begin{align*}
  \denotef{\ctx_1;\ctx}{\heap_1} \Rightarrow \denotef{\many{\tb{x_i}{\ty_i}};\ctx}{\heap'}
  \quad\mathit{and}\quad
  \denotef{\ctx_2;\ctx}{\heap_2} \Rightarrow \denotef{\many{\tb{x_i}{\ty_i}};\ctx}{\heap'}
\end{align*}
 and thus, by consequence,
\begin{align*}
  \htriplefun{\Phi}{e \wedge \denotef{\ctx}{\heap}}{s_1}{\denotef{\many{\tb{x_i}{\ty_i}};\ctx}{\heap'}}\\
\intertext{and}
  \htriplefun{\Phi}{\neg e \wedge \denotef{\ctx}{\heap}}{s_2}{\denotef{\many{\tb{x_i}{\ty_i}};\ctx}{\heap'}}
\end{align*}
hence, by the conditional rule,
\[ \htriplefun{\Phi}{\denotef{\ctx}{\heap}}{\ite{\expr}{s_1}{s_2}}{\denotef{\many{\tb{x_i}{\ty_i}};\ctx}{\heap'}} \]


    \myline
  \item[\textbf{Case} \concBind{x}{z}:] \hfill \\
    Letting $\denotef{\ctx}{\heap} = G \hpjoin H$ by \cref{l:decomp},
by assumption\[
\hassert{
  G \hpjoin H
}
\]
By hypothesis and \cref{l:ptrtype},\[
\hassert{
  G \hpjoin H \hpjoin (x = \loc \wedge \loc \neq null) \hpjoin \shprop{\ty_y}{\loc}{y}
}
\]
By consequence, assuming without loss of generality that $\ty_y$ is a
record type, substitution on the rule for \concName, and applying the
frame rule,
\begin{align*} &\hassert{
G \hpjoin H \hpjoin (x = \loc \wedge \loc \neq null)
\hpjoin \sepmap{x}{y} \hpjoin \sprop{\ty_y}{y}
}\\
&\concBind{x}{z}\\
&\hassert{
G \hpjoin H \hpjoin (x = \loc \wedge \loc \neq null)
\hpjoin \sepmap{x}{y} \hpjoin \sprop{\ty_y}{y} \wedge z = y
},
\end{align*}
since $z$ is a fresh variable.
By definition of $\denote{-}$ and $\ty_{y'}$,
since $z$ does not appear free in $G, H$ or $T_y$,
\[
\hassert{
\denotef{\tybind{y}{\ty_y};\ctx}{\hplocty{\loc}{z}{\ty_{z}}\hpjoin\heap'}
}
\]

    \myline
  \item[\textbf{Case} $\pad{\loc}{x}$] \hfill \\
    Beginning with the proof rule for $\pad{\loc}{x}$,
\begin{align*}
  \htriple{\hpemp}{\pad{\loc}{x}}{\hpemp \wedge \loc = \vnull \wedge x = \vnull}
\end{align*}
By hypothesis, neither $\loc$ nor $x$ appear in $\ctx$ or $\heap$, and thus
do not appear in \denotef{\ctx}{\heap}, so we can apply the frame rule:
\begin{align*}
  \htriple{\denotef{\ctx}{\heap}}{\pad{\loc}{x}}{\denotef{\ctx}{\heap}\hpjoin(\loc = \vnull\wedge x = \vnull)}
\end{align*}
By consequence,
\begin{align*}
  \htriple{\denotef{\ctx}{\heap}}{\pad{\loc}{x}}{\denotef{\ctx}{\heap}\hpjoin\denote{\hplocty{\loc}{x}{\ty}}}
\end{align*}
And thus,
\begin{align*}
  \htriple{\denotef{\ctx}{\heap}}{\pad{\loc}{x}}{\denotef{\ctx}{\heap\hpjoin\hplocty{\loc}{x}{\ty}}}
\end{align*}

    \myline
  \item[\textbf{Case} $\unwindBind{\loc}{\many{z}}$] \hfill \\
    Letting $\denotef{\ctx}{\heap} = G \hpjoin H \hpjoin \denote{\hplocty{\loc}{x}{\tyapp{\tycon}{\many\ty}}}$ by \cref{l:decomp}
and by assumption,
\[
\hassert{
  G \hpjoin H \hpjoin
  (((\loc \neq \vnull) \Rightarrow \shtapp{\many{\ty}}{\loc}{x})
   \wedge ((\loc = \vnull ) \Rightarrow (x = \vnull)))
}
\]
First, assume $\loc = \vnull$.
Then,
\[
\hassert{
  G \hpjoin H \hpjoin (\loc = \vnull \wedge x = \vnull)
}
\]
And by \cref{l:wfmeasures}, 
\[
\hassert{
  G \hpjoin H \hpjoin (x = \vnull) \wedge 
\bigwedge\limits_{\fn{m}} \sepmeas{\fn{m}(null)} = e_{\fn{m}}
}
\]
Quantifying over $x$ allows us to complete the proof via consequence and the
frame rule:
\begin{align*}
&\hassert
{
\denotef{\ctx}{\heap}
}\\
&\hassert
{
\exquant{\many{\fv{...}}}
G \hpjoin H \hpjoin (\loc = \vnull \wedge x = \vnull) \wedge 
\bigwedge\limits_{\fn{m}} \sepmeas{\fn{m}(null)} = e_{\fn{m}}
}\\
&\unwindBind{\loc}{y\cdot\many{y}}
\\
&\hassert{
G \hpjoin H \hpjoin (\loc = \vnull x = \vnull) \wedge 
\bigwedge\limits_{\fn{m}} \sepmeas{\fn{m}(null)} = e_{\fn{m}}
}
\\
&\hassert{
  \denotef{\bigwedge\limits_{\fn{m}} \sepmeas{\fn{m}(y)} = e_{\fn{m}};\ctx}{\heap'}
}
\end{align*}

Next, we assume $\loc \neq \vnull$.
By definition,
\begin{align*}
  \hassert{
G \hpjoin H \hpjoin
\exquant{x_c}
\exquant{\fv{\heap_c}}
\denote{\hplocty{\loc}{x_c}{\ty_c} \hpjoin \heap} 
\wedge \sepunfold{\loc}{x}
}
\end{align*}
By hypothesis, $x_c$ and the free variables of $\denote{\heap_c}$ have
been $\alpha$-renamed and do not appear in $\denote{\ctx}$ or
$\denote{\heap}$.
By the \unwindName axiom and the frame rule, 
\begin{align*}
&
\hassert{
G \hpjoin H \hpjoin
\exquant{x_c}
\exquant{\fv{\heap_c}} 
\denote{\hplocty{\loc}{x_c}{\sub{\many\tvar}{\many\ty}\ty_c}
         \hpjoin \sub{\many\tvar}{\many\ty}\heap_c} 
\wedge \sepunfold{\loc}{x}
}\\
&
{\unwindBind{\loc}{x_c\cdot\dom{\heap_c}\cdot\binder{\heap_c}}}\\
&
\hassert{
G \hpjoin H \hpjoin 
\denote{\hplocty{\loc}{x_c}{\sub{\many\tvar}{\many\ty}\ty_c}
         \hpjoin \sub{\many\tvar}{\many\ty}\heap_c} 
\wedge \sepunfold{\loc}{x}
}
\end{align*}
where $x \in \purety{\tycon}$. 
Therefore, by \cref{l:wfmeasures}, each well-formed measure on
$\tyapp{\tycon}{\tvar}$, $\fn{m}$, is defined on $x$ and we
strengthen:
\begin{align*}
\hassert{&
G \hpjoin H \hpjoin \denote{\hplocty{\loc}{x_c}{\sub{\many\tvar}{\many\ty}{\ty_c}}
         \hpjoin {\sub{\many\tvar}{\many\ty}}\heap_c}
\wedge \sepunfold{\loc}{x}
\wedge \bigwedge\limits_{\fn{m}} \sepmeas{\fn{m}(x)} = e_{\fn{m}}
}
\end{align*}
Which is, by definition
\begin{align*}
\hassert{
  \denote{
\bigwedge\limits_{\fn{m}} \sepmeas{\fn{m}(x)} = e_{\fn{m}};
\ctx} \wedge \denote{\heap'}
}
\end{align*}

    \myline
  \item[\textbf{Case} $x =_z \alloc{\many{\tybind{\field}{e_\field}}}$] \hfill \\
    By assumption, $x$ does not appear bound in $\ctx$ and $\loc$ does not appear in
\denotef{\ctx}{\heap}.
%
%
%
Applying the frame rule to the rule for \allocName gives
us the precondition
\begin{align*}\hassert{
  \denotef{\ctx}{\heap}
}
\end{align*}
and the postcondition
\begin{align*}
  \hassert{
  \denotef{\ctx}{\heap}
  \hpjoin
    \evar = \loc
    \wedge \loc \neq \vnull
    \wedge
    \sepmap{\evar}{z} \wedge z = {\tgenobj{\field}{\expr}}
    }
\end{align*}
Rearranging and by consequence:
\begin{align*}
  \hassert{
  \denotef{\ctx}{\heap}
  \hpjoin
    \evar = \loc
    \wedge \loc \neq \vnull
    \wedge
    (\loc \neq \vnull \Rightarrow
    \sepmap{\loc}{z} \wedge z = {\tgenobj{\field}{\expr}})
  }
\end{align*}
By the typing hypotheses and \cref{l:exprtypelemma}, 
\begin{align*}
  \denote{\ctx} \Rightarrow \sprop{\ty_f}{e}
\end{align*}
\begin{align*}
  \hassert{&
  \denotef{\ctx}{\heap}
  \hpjoin
    \evar = \loc
    \wedge \loc \neq \vnull
    \wedge
    (\loc \neq \vnull \Rightarrow
    \sepmap{\loc}{z} \wedge z = {\tgenobj{\field}{e}} \wedge
     \many{\sprop{\ty_f}{e}} )
     \hfill
  }
\end{align*}
so \cref{a:field} and the definition of \shpropn imply,
\begin{align*}
  \hassert{
  \denotef{\ctx}{\heap\hpjoin\hplocty{\loc}{z}{\ty}}
  }
\end{align*}


    \myline
  \item[\textbf{Case} $y = \access{\evar}{\field_i}$] \hfill \\
By hypothesis, \cref{l:decomp} and \cref{l:ptrtype}, we let 
$$ \denotef{\ctx}{\heap} = G \hpjoin H \hpjoin 
\shprop{\tobj{\ldots \tybind{\field_i}{\objfield{z}{\field_i}} \ldots}}
       {\loc}{z}$$
and by consequence,
\begin{align*}\hassert{
G \hpjoin H \hpjoin \sprop{\tgenobj{\field}{\ty_i}}{z}  
\hpjoin \sepmap{x}{\tobj{\ldots \tybind{\field_i}{\objfield{z}{\field_i}} \ldots}}
}
\end{align*}
By the frame rule:
\begin{align*}&\hassert{
G \hpjoin H \hpjoin \sprop{\tgenobj{\field}{\ty_i}}{z}  
\hpjoin \sepmap{x}{\tobj{\ldots \tybind{\field_i}{\objfield{z}{\field_i}} \ldots}}
}\\
&y = \access{x}{\field_i}\\
&\hassert{
G \hpjoin H \hpjoin \sprop{\tgenobj{\field}{\ty_i}}{z}  
\hpjoin \sepmap{x}{\tobj{\ldots \tybind{\field_i}{\objfield{z}{\field_i}} \ldots}}
\wedge
(y = \objfield{z}{\field_i})
}
\end{align*}
Unfolding the definition of $\spropn$ and by consequence (using the equality
$y = \objfield{z}{\field_i}$),
\begin{align*}
&\hassert{
G \hpjoin H \hpjoin \sprop{\tgenobj{\field}{\ty_i}}{z}  
\hpjoin \sepmap{x}{\tobj{\ldots \tybind{\field_i}{\objfield{z}{\field_i}} \ldots}}
\wedge
\sprop{y}{\ty_i}
}
\end{align*}
And hence, by definition and consequence (to weaken the mapping of $x$
and applying the equality $x = \loc$),
\begin{align*}
\hassert{
  \denotef{\tybind{y}{\ty_i};\ctx}{\heap}
}
\end{align*}
%

    \myline
  \item[\textbf{Case} $\access{\evar}{\field_i} =_z \expr$:] \hfill \\
    Starting from $\hassert{\denotef{\ctx}{\heap}}$,
by \cref{l:decomp,l:ptrtype}, we deduce by consequence,
\begin{align*}
\hassert{
(G &\hpjoin (x = \loc \wedge \loc \neq null) \hpjoin H
\hpjoin \sprop{\tgenobj{\field_j}{\reftp{\tbase_j}{p_j}}}{y} 
\hpjoin \sepmap{x}{\tgenobj{\field_j}{\objfield{y}{\field_j}}})
}
\end{align*}
Thus, by the frame rule,
\begin{flalign*}
\hassert{
(G &\hpjoin (x = \loc \wedge \loc \neq null) \hpjoin H
\hpjoin \sprop{\tgenobj{\field_j}{\reftp{\tbase_j}{p_j}}}{y} 
\hpjoin \sepmap{x}{\tgenobj{\field_j}{\objfield{y}{\field_j}}})
}&
\\
\access{x}{\field_i} & =_z e&
\\
\hassert{
(G &\hpjoin (x = \loc \wedge \loc \neq null) \hpjoin H
\hpjoin \sprop{\tgenobj{\field_j}{\reftp{\tbase_j}{p_j}}}{y} 
\hpjoin \sepmap{x}{z}\wedge z = \tobj{\ldots \tybind{\field_i}{\expr} \ldots})
}&
\end{flalign*}
By consequence and \cref{l:exprtypelemma},
\begin{align*}
\hassert{
&(G \hpjoin \sprop{\reft}{e} \hpjoin (x = \loc \wedge \loc \neq null) \hpjoin H \\
&{ }\hpjoin \sprop{\tgenobj{\field_j}{\reftp{\tbase_j}{p_j}}}{y} 
\hpjoin \sepmap{x}{z}\wedge z = \tobj{\ldots \tybind{\field_i}{\expr} \ldots})
}
\end{align*}
Unfolding the definition of \spropn, this expands to
\begin{align*}
\hassert{
&(G \hpjoin \sprop{\reft}{e} \hpjoin (x = \loc \wedge \loc \neq null) \hpjoin H \\
&{ }\hpjoin 
y = \tgenobj{\field_j}{\objfield{y}{\field_j}} 
\wedge
\bigwedge_{\mathclap{\tybind{\field_j}{\ty_j}}}
\sprop{\ty_j}{\objfield{y}{\field_j}}
\\
&{ }\hpjoin
\sprop{\tgenobj{\field_j}{\reftp{\tbase_j}{p_j}}}{y} 
\hpjoin \sepmap{x}{z}\wedge z = \tobj{\ldots \tybind{\field_i}{\expr} \ldots})
}
\end{align*}
By \cref{a:field},
\begin{align*}
\hassert{
&(G \hpjoin \sprop{\reft}{e} \hpjoin (x = \loc \wedge \loc \neq null) \hpjoin H \\
&{ }\hpjoin 
y = \tgenobj{\field_j}{\objfield{y}{\field_j}} 
\wedge
\bigwedge_{\mathclap{\tybind{\field_j}{\ty_j}}}
\sprop{\ty_j}{\objfield{y}{\field_j}}
\\
&{ }\hpjoin
\sprop{\tgenobj{\field_j}{\reftp{\tbase_j}{p_j}}}{y} 
\hpjoin \sepmap{x}{z}
\\
&{ }
\wedge z = \tobj{\tybind{\field_0}{\objfield{y}{\field_0} \ldots \tybind{\field_i}{\objfield{z}{\field_i}} \ldots}}\\
&{ }\wedge (\objfield{z}{\field_i} = e)
}
\end{align*}
By definition of \shpropn,
\begin{align*}
\hassert{
&(G \hpjoin \sprop{\reft}{e} \hpjoin (x = \loc \wedge \loc \neq null) \hpjoin H \\
&{ }\hpjoin 
y = \tgenobj{\field_j}{\objfield{y}{\field_j}} 
\wedge
\bigwedge_{\mathclap{\tybind{\field_j}{\ty_j}}}
\sprop{\ty_j}{\objfield{y}{\field_j}}
\\
&{ }\hpjoin
\sprop{\tgenobj{\field_j}{\reftp{\tbase_j}{p_j}}}{y} 
\hpjoin 
\shprop{\ty_r}{\loc}{z}
}
\end{align*}
and thus, by definition,
\begin{align*}
&\hassert{
  \denotef{\ctx}{\heap}
}\\
&\access{x}{\field_i} = e
\\
&\hassert{
  \denotef{\ctx}{\hplocty{\loc}{z}{\ty_r}\hpjoin\heap}
}
\end{align*}

    \myline
  \item[\textbf{Case} $\windBind{\loc}{z}$] \hfill \\
Given
$ \tydef{\tyapp{\tycon}{\tvar}}
                         {\heap}
                         {\evar}
                         {\ty}$,
\[
  \shtapp{\ty_{\tvar}}{\loc}{y} \myeq
    (\exquant{x,\fv{\heap}}
      \denote{\hplocty{\loc}{x}{\sub{\tvar}{\ty_{\tvar}}\ty}\hpjoin\sub{\tvar}{\ty_{\tvar}}\heap}
    \wedge 
    \sepunfold{\loc}{y} 
\]
By \cref{l:folding} applied to the type judgement folding assumption we deduce
\[
\exquant{\fv{\heap_c}}
\denote{\hplocty{\loc}{x}{\theta\ty_c} \hpjoin \theta\heap_c}
\wedge 
\exquant{y}\sepunfold{\loc}{y}
\quad
\mbox{where } 
\quad
\theta = \many{\sub{\tvar}{\ty}}
\]
and, by the definition of \shtapp{\many\ty}{\loc}{x}:
\begin{align*}
  \exquant{y}\shtapp{\many\ty}{\loc}{y}
\end{align*}
Applied to the rule for \windBind{\loc}{y}, since by assumption $y$ is
a fresh variable,
\begin{align*}
\htriple{\exquant{y}\shtapp{\many\ty}{\loc}{y}}
        {\windBind{\loc}{y}}
        {\shtapp{\many\ty}{\loc}{y}}
\end{align*}
By \cref{l:pure}, $x \in \purety{\tyapp\tycon{\many{\ty}}}$, so by
\cref{l:wfmeasures}, each well-formed measure on
$\tyapp{\tycon}{\tvar}$, $\fn{m}$, is defined on $y$.
By the form of $\ty_c$ and the definition of \sepunfoldN,
for any field of $x$
$\objfield{y}{\field} = \objfield{x}{\field}$
and we 
thus strengthen the postcondition:
\[
          \hassert{\shtapp{\many\ty}{\loc}{y}
            \wedge \bigwedge\limits_{\fn{m}}\fn{m}(y) = e_{\fn{m}}}.
\]
Applying the frame rule and consequence with the definition of $\ty_y$
and \cref{l:subtyping} applied to $\heap'$,
\begin{align*}
  \htriplefun{\Phi}
  {\denotef{\ctx}{\hplocty{\loc}{x}{\ty_x}\hpjoin\heap_x\hpjoin\heap}}
  {\windBind{\loc}{y}}
  {\denotef{\ctx}{\hplocty{\loc}{y}{\ty_y}\hpjoin\heap'}}
\end{align*}

    \myline
  \item[\textbf{Case} $x_r = \fun(\many{x})$] \hfill \\
    Assuming 
$
\tybind{\fun}{\funschema}$ and
\[
\funschema = 
{\quant{\many{\loc}}\quant{\many{\tvar}}\funty{\many{\tybind{\evar}{\ty}}}{\heap_i\hpjoin\heap_i'}{\equant{\many{\loc_o}}\tybind{\evar_o}{\ty_o}}{\heap_o}}
\]
%
let 
$P = \Pre{\funschema}$
and 
$Q = \Post{\funschema}$.
We apply the substitution $\theta = \many{\sub{x_j}{\expr_j}}$ to obtain:
\begin{align*}
\theta P &= \denote{\many{\tybind{e_j}{\theta\ty_j}}}\wedge\denote{\theta\heap_i}\\
\theta Q &= \denote{{\tybind{x_o}{\theta\ty_o}}}\wedge\denote{\theta\heap_o}
\end{align*}
By \cref{l:decomp,l:subtyping},
\begin{align*} 
  \denote{\ctx} & \Rightarrow \many{\sprop{\theta\ty_j}{e_j}}\\
  \denote{\ctx} & \Rightarrow \denote{\heap_m} \Rightarrow \denote{\theta \heap_i}
\end{align*}
and thus
\begin{align*}
  \denote{\ctx} & \Rightarrow \denote{\heap_u\hpjoin\heap_m} \Rightarrow \denote{\ctx} \wedge \theta{P} \hpjoin \heap_u
\end{align*}
Which gives us, by consequence, and framing of $\ctx$ and $\heap_u$, and unfolding the definition of $Q$,
\[
\htriplefun{\Phi}
           {\denotef{\ctx}{\heap_m\hpjoin\heap_u}}
           {x = f(\many{e_j})}
           {\denotef{\ctx}{\heap_u\hpjoin\theta\heap_o}}
\]

\end{enumerate}  
\qed
\end{proof}


\showprocthm
\begin{proof}
  Let
\begin{align*}
P &= \Pre{\funschema} = \denote{\many{\tybind{x}{\ty}}}\wedge\denote{\heap_i}&\\
Q &= \Post{\funschema} = \denote{{\tybind{x_o}{\ty_o}}}\wedge\denote{\heap_o}&
\end{align*}
By inversion on the judgement, we must have typed a
statement $s' \doteq \return{e}$ for some expression, or a sequence
of statements ending in a \returnName.
By the statement typing theorem,
\[
\htriplefun{\Phi}
{\Pre{\funschema}}
{s}
{\denotef{\ctx_s}{\heap_s}}
\]
Applying \cref{l:exprtypelemma} and \cref{l:subtyping} to the
hypotheses of the \returnName typing rule, we determine
\begin{align*}
\denotef{\ctx_s}{\heap_s} & \Rightarrow \sub{x_o}{e}\sprop{\ty_o}{x_o}\\
\denotef{\ctx_s}{\heap_s} & \Rightarrow \sub{x_o}{e}\denote{\heap_o}.
\end{align*}
and thus
\[
\htriplefun{\Phi}
{\denotef{\ctx_s}{\heap_s}}
{\return{e}}
{\Post{\funschema}}.
\]
By the sequence rule,
\[
\htriplefun{\Phi}
{\Pre{\funschema}}
{s;\return{e}}
{\Post{\funschema}}.
\]
\qed

\end{proof}

\showtypeproofs

\FloatBarrier
\section{Heap Annotation Inference}\label{sec:annots}

Up until this point, our presentation has assumed that the
annotations \concName, \unwindName, and \windName have already
been inserted into the source file.
To alleviate this burden, \toolname automatically inserts these
annotations at critical locations in the input program.
To be sure, the soundness of the type system does not depend on
any particular algorithm for inferring these locations.
In fact, while the algorithm we will present was sufficient for the
benchmarks we used to test \toolname, it would be entirely feasible to
swap our algorithm for another.
In this section, we will describe the simple source elaboration that
preceeds refinement type checking and inference.

\mypara{Annotation inference.}
To formalize our method of inferring annotation locations, we
define a function $\exformName$ with type:
\[
\exformName: \textsf{Statement} \times \exformStateVar \rightarrow \textsf{Statement} \times \mathcal{P}(\exformUnwound)
\]
\begin{align*}
\exformStateVar & \myeq \textsf{Type Bindings} \times \textsf{Heap} \times \textsf{Function} \times \exformUnwound\\
\exformUnwound & \myeq \textsf{Location} \times \textsf{Type Constructor}
\end{align*}
Thus, $\exformEq{\stmt}$ transforms the statement $\stmt$ yielding a
new (possibly compound, \ie comprising several sequenced statements)
statement and a new set of unfolded locations.
The input to $\exformName$ is
$\stmt$, the statement to be transformed;
$\bctx$, containing physical type information for local variables;
$\bheap$, containing physical heap type information;
$\fun$, the physical specification of the function containing $\stmt$; and
$\exformUnwound$, a set of $(\loc,\tycon)$ pairs that denote unfolded locations
and the type constructor that was unfolded.

Because this step \emph{preceeds} refinement type checking, the only
type information that is available is \emph{base} type information -- \ie 
all types are of the form $\reftp{\tbase}{true}$.
Definitions for $\exformName$ on the statements that may possibly
generate annotations are given in \cref{fig:annotations}, with
some helper functions given in \cref{fig:annotation-helpers}.

\mypara{Function Declarations.}
$\exformName$ transforms programs function by function.
Thus it simply calls $\exformName$ recursively on the function body,
querying the type system to determine $\bctx$ and $\bheap$ with
$\fn{Env}(\fun)$ and $\fn{HeapIn}(\fun)$.

\begin{figure*}[t]
\begin{flalign*}
  \exformEqState{\fundecl{\fun}{\many{\evar}}}{\exformStateVar} & = 
    (\fundeclstmt{\fun}{\many{\evar}}{s'}, \emptyset)&\\
    \mbox{\textit{where }} 
      (s',-) & = \exformEqState{s}{\mkexformState{\textsf{Env}(\fun)}
                                         {\textsf{HeapIn}(\fun)}
                                         {\fun}
                                         {\emptyset}}&\\
                                       &\\
\exformEq{y = \access{\evar}{\field_i}} &= 
   (\conc{\evar};u;y = \access{\evar}{\field_i},\exformUnwound') &
   &\\
   \mbox{\textit{where }} 
   (u, \exformUnwound') &= \unwindFun{\evar,\bctx,\bheap,\exformUnwound}&\\
   &\\
\exformEq{\access{\evar}{\field_i} = \expr} &= 
   (\conc{\evar};u;\access{\evar}{\field_i}=\expr;\conc{\evar},\exformUnwound') &
   &\\
\mbox{\textit{where }}
   (u, \exformUnwound') &= 
     \unwindFun{\evar,\bctx,\bheap,\exformUnwound}&\\
     &\\
\exformEq{\evar = \funcallalt{\expr}} & = 
   (w;p;\evar = \funcallalt{\expr},\exformUnwound') &\\
\mbox{\textit{where }} 
   p & = \padlocs{\heap,\heapIn{\funalt}}&\\
  (w,\exformUnwound') & = \windFun{\heap,\heapIn{\funalt}, \exformUnwound}&\\
  &\\
  \exformEq{\return{\expr}} & =
  (w;p;\return{\expr},\emptyset)&\\
\mbox{\textit{where }} 
   p & = \padlocs{\heap,\heapOut{\fun}}&\\
  (w,\exformUnwound') & = \windFun{\heap,\heapOut{\fun}, \exformUnwound}&\\
  &\\
\exformEqState{\ite{\expr}{\stmt_1}{\stmt_2}}{\exformState\mbox{ as }\exformStateVar} & =
  (\ite{\expr}{\stmt_1';w_1;p_1}{\stmt_2';w_2;p_2},\exformUnwound \setminus L_{alias}) &\\
\mbox{\textit{where }}
  p_1 & = \padlocs{\heap_1, \heap_2}&\\
  p_2 & = \padlocs{\heap_2, \heap_1}&\\
  (\stmt_1',\exformUnwound_1) & = \exformEqState{\stmt_1}{\exformStateVar}&\\
  (\stmt_2',\exformUnwound_2) & = \exformEqState{\stmt_2}{\exformStateVar}&\\
  (\heap_1, \heap_2) & = (\heapAfter{\stmt_1}, \heapAfter{\stmt_2})       &\\
  L_{alias} & = 
   \{ (\loc,\tycon) \mid 
    \fn{Alias}(\loc,\heap_1,\heap_2) \wedge (\loc, C) \in \exformUnwound \}&\\
  (L_1, L_2) & = (\exformUnwound_1 \setminus \exformUnwound, \exformUnwound_2 \setminus \exformUnwound) &\\
  w_1    & = \topsort(\{ \wind{\loc} \mid (\loc,\tycon) \in L_1 \cup L_{alias} \},\bheap)&\\
  w_2    & = \topsort(\{ \wind{\loc} \mid (\loc,\tycon) \in L_2 \cup L_{alias} \},\bheap)
\end{flalign*}
\caption{Statement annotation insertion}
\label{fig:annotations}
\end{figure*}
\begin{figure*}
  \centering
\begin{flalign*}
\unwindFun{\evar, \bctx, \bheap, \exformUnwound} & = (U,L\cup\exformUnwound) &\\
   \mbox{\textit{where }}
    U & = \{ \unwind{\loc} \mid (\loc, \tycon) \in L\setminus\exformUnwound \}&\\
    L & = \{ (\loc,\tycon) \mid 
             \loc \in \locquery{\bctx}{\evar}
             \wedge (\hplocty{\loc}{y}{\tyapp{\tycon}{\many{\tbase}}}) \in \bheap
                       \}&
\end{flalign*}
\eqsep
\begin{flalign*}
\windFun{\heap, \heap', \exformUnwound} & = (W ,L \cap \exformUnwound)&\\
\mbox{\textit{where }}
 W & = \topsort(\{ \wind{\loc} \mid (\loc,\tycon) \in L' \setminus L \}, \heap) \\
 (L, L') & = (\woundLocs{\heap}, \woundLocs{\heap'}) &
&
\end{flalign*}
\eqsep
\begin{flalign*}
  \fn{Alias}(\loc, \heap_1, \heap_2) & =
   \hplocty{\loc}{x_1}{\tobj{\ldots\tybind{\field}{\tbase_1}\ldots}} \in \heap_1
  \wedge
   \hplocty{\loc}{x_2}{\tobj{\ldots\tybind{\field}{\tbase_2}\ldots}} \in \heap_2   &\\ 
   & \quad \wedge
   |\locs{\tbase_1} \cup \locs{\tbase_2}| > 1&
&
\end{flalign*}
\eqsep
\begin{flalign*}
\locquery{\bctx}{\evar} & = \{ \locs{\tbase} \mid \hastype{\bctx}{\evar}{\tbase} \}&
\end{flalign*}
\eqsep
\begin{flalign*}
\woundLocs{\heap} & = \{ (\loc, \tycon) \mid \hplocty{\loc}{\evar}{\tyapp{\tycon}{-}} \in \heap \}&
\end{flalign*}
\begin{flalign*}
\padlocs{\heap,\heap'} & = \{ \padnone{\loc} \mod \loc \in \dom{\heap'} \setminus \dom{\heap} \}&
\end{flalign*}
\caption{Annotation insertion helper functions}
\label{fig:annotation-helpers}
\end{figure*}



%

\mypara{Field Access and Mutation.}
On a field read or write, \toolname computes the set of locations
that must be unfolded, using the \fn{UnfoldList} helper.
\fn{UnfoldList} computes this list by
\begin{inparaenum}[(1)]
\item consulting $\ctx$ and $heap$ to determine if $x$ points to a location
that is folded up; and
\item subtracting from this list all locations that are already unfolded
\end{inparaenum}.
\toolname inserts \unwindName calls for these locations before the
field access.

\toolname also optimistically inserts a \concName before a field access.
The \concName will type check exactly when the field access type checks.

\mypara{Function Return.}
At function returns, \toolname must take care to ensure that $\bheap$
agrees with the current function's specified heap with respect to
which locations folded.
If a location in the function's schema is unfolded, it must first be folded up.
At a return statement in a function, $\fun$, \toolname computes these
locations by comparing the current heap, $\heap$ with $\fun$'s
specified heap.
Given a ``current'' heap $\heap$ and a ``target'' heap $\heap'$,
$\fn{FoldList}$ computes the set of locations in $\heap$ that must be
folded up.
These locations may depend on each other, so $\fn{FoldList}$ orders
these \windName calls using the $\fn{FoldOrder}$ function.

\mypara{Function Calls.}
With respect to \windName{s}, function calls behave exactly like function
returns.

\myexsh.
Revisiting \code{abslist}, consider the following:
\begin{lstlisting}
var l1 = { data:0, next:null };
var l2 = { data:1, next:l1   };
absList(l2);
\end{lstlisting}
$\fn{FoldList}$ determines that the locations (and their associated
types) that must be folded up before calling \code{abslist} are
(\code{\&l2}, \code{List[number]}) and (\code{\&l1}, \code{List[number]}).
However, \code{\&l2} depends on \code{\&l1}.
$\fn{FoldOrder}$ realizes this dependency with the correct ordering
\code{fold(\&l1); fold(\&l2);}.
The code is annotated with \windName and \concName calls are as follows:
\begin{src}
var l1 = { data:0, next:null };
var l2 = { data:1, next:l1   };
//: fold(&l1);
//: fold(&l2);
absList(l2);
\end{src}

\mypara{If Statements.} 
To annotate \code{if} statements, \toolname calls $\exformName$
on the statements in the ``then'' and ``else'' branches.
To determine which statements must be folded, $\exformName$ calculates
three sets.
$L_1$ and $L_2$ are the sets of unfolded locations that were previously folded \emph{before} the \code{if} statement.
Additionally, $\exformName$ queries the type checker to get heaps
$\heap_1$ and $\heap_2$ that are returned from type checking $\stmt_1$
and $\stmt_2$, respectively.
These heaps are used to determine $L_{alias}$, by calling \fn{Alias}. 
This procedure determines if a location $\loc$ would cause any references to
become aliased (and thus eventually fail to type check).
Using the same ordering as in function returns and function calls, 
\toolname calls $\topsort$ to determine the sequence of folds that
need to occur in both the ``then'' and ``else'' branches.

\myexsh.
In the following, suppose @x@ is a pointer to a list.
The code
\begin{src}
var d = x.data;
if (d > 0) {
  x.next = { data: 1, next: null };
} else {
  x.next = { data: -1, next: null };
}
\end{src}
would thus be annotated
\begin{src}
//: unfold(&x)
var d = x.data;
if (d > 0) {
  x.next = { data: 1, next: null };
  //: fold(&x)
} else {
  x.next = { data: -1, next: null };
  //: fold(&x)
}
\end{src}
because the \code{next} fields of \code{x.next} at the end of either
branch point to \emph{different} locations.
Folding $\&x$ ensures a consistent view of the heap after the control
flow join.

\mypara{Padding}
Whenever a heap subtyping occurs, it is possible that the sub-heap has
fewer locations than the super-heap.
However, the heap subtyping judgements requires the sub-heap and super-heap
to have equivalent domains.
We insert \padName statements at these locations when the
physical type checker determines that the sub-heap's domain is too
small.



\FloatBarrier


\else
\fi

\end{document}